\documentclass[onecolumn,aps,floatfix,showkeys,superscriptaddress]{revtex4-2}

\usepackage[cmex10]{amsmath}
\usepackage{amsthm}
\usepackage{amssymb}
\usepackage{amsfonts}
\usepackage{amstext}
\usepackage{mathtools}
\usepackage{IEEEtrantools}

\usepackage{array}
\usepackage{tikz}
\usepackage{pgfplots}
\pgfplotsset{compat=newest} 
\pgfplotsset{plot coordinates/math parser=false}
\usetikzlibrary{quantikz}
\usepackage{graphicx}
\usepackage{rotating}
\usepackage{subcaption}

\usepackage{bibentry}

\usetikzlibrary{decorations.markings,positioning,calc}

\usepackage{xifthen,xparse}
\usepackage{stmaryrd}

\usepackage{color, soul, colortbl}
\definecolor{darkgreen}{rgb}{0,0.5,0}
\definecolor{magenta}{rgb}{1,0,1}
\definecolor{frankfurt_blue_body}{rgb}{0.2,0.2,0.698}

\usepackage{float}

\usepackage{etoolbox}
\let\bbordermatrix\bordermatrix
\patchcmd{\bbordermatrix}{8.75}{4.75}{}{}

\usepackage{textcomp}

\allowdisplaybreaks

\newtheorem{remark}{Remark}
\newtheorem{lemma}{Lemma}
\newtheorem{example}{Example}

\newenvironment{example*}
  {\addtocounter{example}{-1}\example}
  {\endexample}

\newcommand{\MCC}{\mathcal{C}}

\newcommand{\vecnot}[1]{\underline{#1}}

\newcommand{\cnot}[2]{{\text{CNOT}}_{#1 \rightarrow #2}}

\newcommand{\tr}[1]{\text{Tr}\left[ #1 \right]}
\newcommand{\norm}[1]{\left\| #1 \right\|}	
\newcommand{\dket}[1]{\left\lvert #1 \right\rangle}

\NewDocumentCommand\dketbra{+m+g}{%
  \IfNoValueTF{#2}
    {\left\lvert #1 \right\rangle \left\langle #1 \right\vert}
  {\left\lvert #1 \right\rangle \left\langle #2 \right\rvert}%
}
\NewDocumentCommand\dbraket{+m+g}{%
  \IfNoValueTF{#2}
    {\left\langle #1 \vert #1 \right\rangle}
  {\left\langle #1 \vert #2 \right\rangle}%
}

\makeatletter
\renewcommand*\env@matrix[1][\arraystretch]{%
  \edef\arraystretch{#1}%
  \hskip -\arraycolsep
  \let\@ifnextchar\new@ifnextchar
  \array{*\c@MaxMatrixCols c}}
\makeatother

\makeatletter
\newcommand{\settitle}{\@maketitle}
\makeatother

\begin{document}

\title{Belief Propagation with Quantum Messages for\\ Quantum-Enhanced Classical Communications}
\author{Narayanan Rengaswamy}
\email{narayananr@arizona.edu, corresponding author; most of this work was done when he was with the Department of Electrical and Computer Engineering, Duke University, Durham, North Carolina 27708, USA}
\affiliation{Department of Electrical and Computer Engineering, University of Arizona, Tucson, Arizona 85721, USA}
\author{Kaushik P. Seshadreesan}
 \email{kaushiksesh@email.arizona.edu, corresponding author}
\affiliation{College of Optical Sciences, University of Arizona, Tucson, Arizona 85721, USA}
\author{Saikat Guha}
 \email{saikat@optics.arizona.edu}
\affiliation{College of Optical Sciences, University of Arizona, Tucson, Arizona 85721, USA}
\author{Henry D. Pfister}
 \email{henry.pfister@duke.edu}
\affiliation{Department of Electrical and Computer Engineering, Duke University, Durham, North Carolina 27708, USA}

\date{\today}

\begin{abstract}

For space-based laser communications, when the mean photon number per received optical pulse is much smaller than one, there is a large gap between communications capacity achievable with a receiver that performs individual pulse-by-pulse detection, and the quantum-optimal ``joint-detection receiver'' that acts collectively on long codeword-blocks of modulated pulses; an effect often termed ``superadditive capacity''.  
In this paper, we consider the simplest scenario where a large superadditive capacity is known: a pure-loss channel with a coherent-state binary phase-shift keyed (BPSK) modulation. 
The two BPSK states can be mapped conceptually to two non-orthogonal states of a qubit, described by an inner product that is a function of the mean photon number per pulse. 
Using this map, we derive an explicit construction of the quantum circuit of a joint-detection receiver based on a recent idea of ``belief-propagation with quantum messages'' (BPQM). 
We quantify its performance improvement over the Dolinar receiver that performs optimal pulse-by-pulse detection, which represents the best ``classical'' approach.
We analyze the scheme rigorously and show that it achieves the quantum limit of minimum average error probability in discriminating $8$ (BPSK) codewords of a length-$5$ binary linear code with a tree factor graph.
Our result suggests that a BPQM-receiver might attain the Holevo capacity of this BPSK-modulated pure-loss channel. Moreover, our receiver circuit provides an alternative proposal for a quantum supremacy experiment, targeted at a specific application that can potentially be implemented on a small, special-purpose, photonic quantum computer capable of performing cat-basis universal qubit logic.

\end{abstract}


\maketitle


\section{Introduction}
\label{sec:intro}


``Message-passing" algorithms are used to efficiently evaluate quantities of interest in problems defined on graphs. 
They work by passing messages between nodes of the graph. 
For example, these algorithms have been successfully used for statistical inference, optimization, constraint-satisfaction problems and the graph isomorphism problem among several other applications~\cite{Yedidia-03,Yedidia-it05,Globerson-nips08,Lu-aller08,Donoho-itw10,Bayati-it11,Yedidia-jsp11,Mansour-arxiv17}. 
In particular, ``belief-propagation'' (BP) is a message-passing algorithm for efficiently marginalizing joint probability density functions in statistical inference problems. 
The algorithm derives its name from the fact that the messages used in BP are ``local'' probabilities or ``beliefs'' (e.g., of the value of the final quantity of interest). 
An important application of BP lies in the decoding of linear codes using the posterior bit-wise marginals given the outputs of a classical channel~\cite{RU-2008}.
It is well-known that BP exactly performs the task of optimal bit-wise maximum-a-posteriori (bit-MAP) decoding when the code's factor graph is a tree.
However, since codes with tree factor graphs have poor minimum distance~\cite{RU-2008}, BP is also applied to codes whose factor graphs have cycles, e.g, low-density parity-check (LDPC) codes. 
Although BP does not compute the exact marginals in this case, it is computationally more efficient than MAP, and usually performs quite well. 
In fact, it has been proven that, for large blocks, BP achieves the optimal MAP performance for spatially-coupled LDPC codes over the binary erasure channel~\cite{Kudekar-it11} and binary memoryless symmetric channels~\cite{Kudekar-it13,Kumar-it14}.
From a more practical perspective, BP-based decoders are routinely deployed in modern communications and data storage.


Given the success of BP decoding for classical channels, it is natural to ask if it can be generalized to the quantum setting.
For example, can one decode classical codes for communications over a classical-quantum channel or, more generally, perform efficient inference on graphically-represented classical data encoded in qubits?
Consider laser communications based on binary-phase-shift-keying (BPSK) modulation for sending classical data over a pure-loss bosonic channel of transmissivity $\eta \in (0,1]$~\cite{Guha-isit12}. 
During each "use" of the quantum channel, the transmitter modulates each optical pulse, or mode, into one of the two coherent states $\dket{\alpha}$ or $\dket{-\alpha}$, where $\alpha \in {\mathbb R}$ and the mean photon number per mode equals $N_S = |\alpha|^2$. 
Each channel output symbol is an optical pulse that is in one of the two coherent states $\dket{\pm \beta}$, where $\beta = \sqrt{\eta} \alpha$ and mean photon number $N = \eta N_S$. 
These two states are non-orthogonal with an inner product $\langle \beta | -\beta \rangle = e^{-2N} \equiv \sigma$. 
In this case, the coherent states $\dket{\pm \beta} = \sum_{n=0}^\infty e^{-|\beta|^2/2}\frac{(\pm\beta)^n}{\sqrt{n!}} \dket{n}$ live in an infinite-dimensional Hilbert space spanned by the complete orthonormal number basis $\left\{ \dket{n}, n \in \mathbb{N} \right\}$.
However, since each channel output is always in one $\dket{\pm \beta}$, for the purposes of designing a receiver, we can embed the subspace spanned by $\dket{\pm \beta}$ in a two-dimensional (qubit) Hilbert space via the inner-product-preserving map: 
\begin{align}
\dket{\pm\beta} \longmapsto \dket{\pm \theta} \coloneqq {\rm cos}\frac{\theta}{2} \dket{0} \pm {\rm sin}\frac{\theta}{2} \dket{1}, 
\end{align}
with $\sigma = {\rm cos} \theta$. 
The resulting channel from a classical encoding variable $x$ to a conditional quantum state, i.e., $[x = 0] \mapsto \dket{\theta}, [x=1] \mapsto \dket{-\theta}$, is often called a (pure-state)  ``classical-quantum'' (CQ) channel in the quantum information theory literature.

When the channel output symbols are detected one at a time, the best possible detection error probability is given by the Helstrom bound~\cite{Helstrom-jsp69,Helstrom-ieee70} on the minimum average error probability of discriminating the alphabet states $\dket{\pm\beta}$, which is $p \coloneqq \frac12[1-\sqrt{1-\sigma^2}]$. 
A structured optical design of a receiver that achieves this performance was invented by Dolinar in 1973~\cite{Dolinar-1973}. 
This receiver induces a binary symmetric channel (BSC) between the quantum channel outputs $\dket{\pm \beta}$ and the receiver's guess $``\pm \beta"$, with crossover probability $p$, thereby enabling the communicating parties to achieve a reliable communication rate given by $C_1 = 1- h_2(p)$ bits per mode, the Shannon capacity of the BSC. 
To achieve communication at a rate close to this capacity, one would need to use a code that achieves the Shannon capacity of the BSC, e.g., Arikan's polar code~\cite{Arikan-it09}, and a suitable decoder. 
If the receiver detects, i.e., converts from the quantum (optical) to the electrical domain, each quantum channel output one at a time, no amount of classical post-processing, including feedforward between channel uses, and soft-information processing, can achieve a rate higher than $C_1$. 
Thus, a capacity-approaching LDPC code for the BSC and a BP decoder can approach but not surpass the rate $C_1$.

However, if one employs a quantum joint-detection receiver that collectively measures the entire block of $n$ channel outputs, then the rate may increase to the Holevo limit, $C_\infty = S(\frac12 \dketbra{\beta} + \frac12 \dketbra{-\beta}) = h_2([1+\sigma]/2)$ bits per mode, where $S(\cdot)$ denotes the von Neumann entropy. 
In the limit as $N \to 0$ (or equivalently $\sigma \to 1$), where the mean photon number per mode vanishes, one can show that  $C_\infty / C_1 \to \infty$ and collective measurement is preferable. 
This regime of operation is especially important for long-haul free-space terrestrial and deep-space laser communications.
In order to fully exploit this large capacity gain, one can use a CQ polar code~\cite{Guha-isit12} with a decoder based on collective measurement of the received quantum state.
Alternatively, one can use a codebook comprising $M = 2^{nR}$ random length-$n$ codewords with $R < C_\infty$, where each symbol of each codeword chosen from an equal prior over the two BPSK symbols.
If the receiver employs a joint measurement that discriminates between the codewords sufficiently well, then the probability of decoding error will converge to $0$ as $n \to \infty$. 
Both the optimal measurement and the square-root measurement (SRM) are known to faithfully discriminate between roughly $2^{nC_\infty}$ codewords~\cite{Wilde-2013}, as opposed to only $2^{nC_1}$ codewords if symbol-by-symbol detection is combined with classical decoding. 
Given the quantum states of the $M$ codewords, the optimal measurement can be computed by applying the Yuen-Kennedy-Lax (YKL) conditions~\cite{Yuen-it75} applied to the Gram matrix of the codebook---the $M$-by-$M$ matrix of pairwise inner products of the codewords' quantum states. 
This calculation is simpler for linear codes~\cite{Eldar-it00}, especially codes that have certain group symmetries~\cite{Krovi-pra15}. 
Even when it is possible to compute the optimal measurement for the codebook, it may be hard to translate the mathematical description into a physical receiver design~\cite{Barthel-prl18}, unless we have a general-purpose photonic quantum computer~\cite{DaSilva-pra13}. 
Therefore, an efficient and physically-realizable receiver is of significant practical interest if it can outperform the optimal receiver based on optimal symbol-by-symbol measurement.


Renes~\cite{Renes-njp17} recently proposed a quantum generalization of BP for a binary-output pure-state CQ channel. 
Renes' algorithm is well-defined on a tree factor graph and works by passing quantum messages (encoded in qubits) and classical messages (bits) between nodes of the code's factor graph. 
Unlike earlier algorithms termed ``quantum belief-propagation''~\cite{Hastings-prb07,Leifer-annphy08}, Renes' algorithm does not measure the $n$ channel outputs, followed by classical BP on the (classical) syndrome measurements, and hence is not limited to achieving a rate of $C_1$. 
In order to avoid any confusion with previous quantum belief-propagation algorithms, we will refer to Renes' algorithm as ``belief-propagation with quantum messages'' (BPQM).

In~\cite{Renes-njp17}, the first step in developing BPQM was to interpret the message-combining operations in classical BP as ``channel combining'' rules that execute a local inference procedure. 
This step has close connections with the channel combining operation defined by Arikan for polar codes~\cite{Arikan-it09}. 
The second step was a generalization of these channel combining rules to allow for quantum messages, as in CQ polar codes~\cite{Wilde-2013}, i.e., messages that are qubit density matrices which capture the node's belief about a message bit. 
The above rules define a CQ channel that gets induced at each node when (quantum) messages arrive at it.
Finally, the third step was to define appropriate unitary operations at the nodes, which process the outputs of the aforesaid induced CQ channels and produce messages to be passed on to subsequent nodes.  

While~\cite{Renes-njp17} is a breakthrough paper that provides a phenomenological description of BPQM, it does not assess its performance on an example code or make comparisons with optimal symbol-by-symbol measurements. 
Also, it does not present a proof of decoding optimality, even for CQ codes with a tree factor graph. 
Finally, it does not prescribe an explicit quantum circuit for BPQM.
This paper addresses all of the above open questions and resolves many of them for the chosen example code.

\section{Results and Discussion}


\subsection{BPQM-based Receiver Design and Block Error Rates}
\begin{figure}
\begin{center}

\includegraphics[scale=1,keepaspectratio]{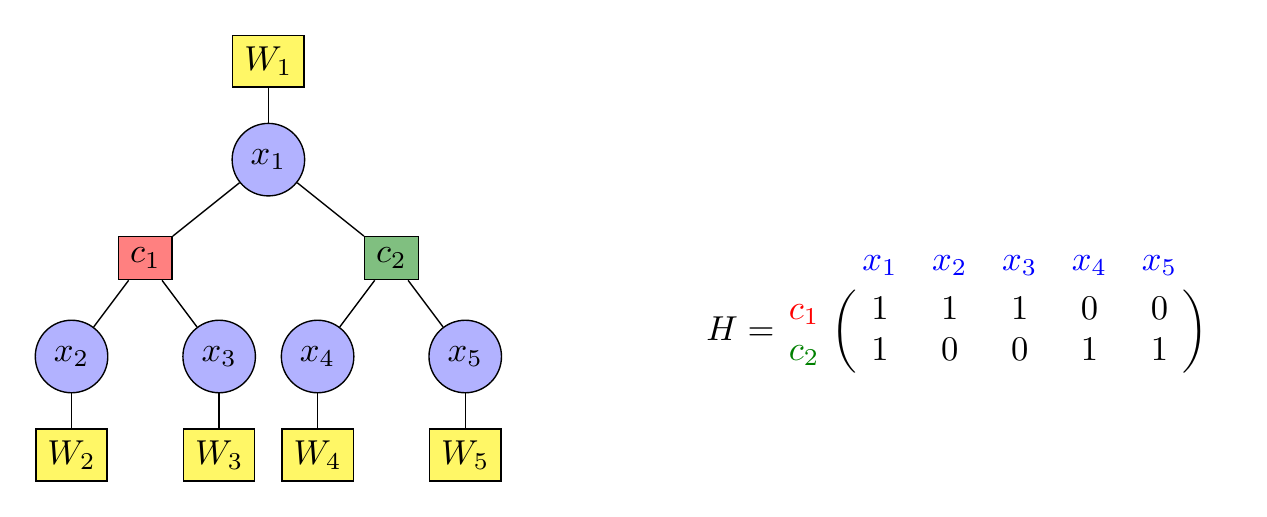}

\caption{\label{fig:five_bit_code}Factor graph and parity-check matrix for the $5$-bit linear code in the running example.}

\end{center}
\end{figure}

\begin{figure}

    \centering
	\hspace{-0.75cm}\includegraphics[scale=0.75,keepaspectratio]{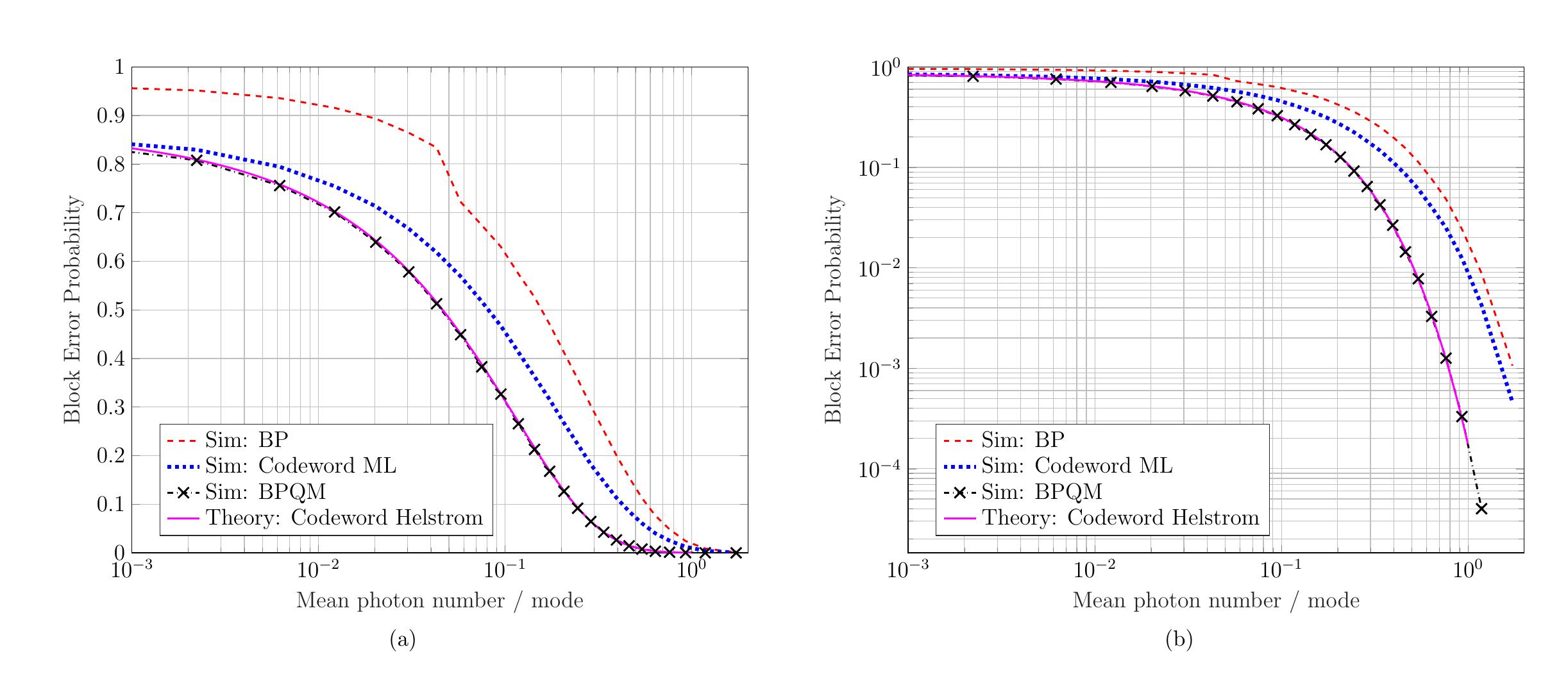}

    \caption{\label{fig:bpqm_perr_vs_nbar} The overall performance of BPQM compared with that of other related schemes. (a) The overall block error rate of BPQM along with those of optimal joint Helstrom, symbol-by-symbol Helstrom followed by classical optimal block-MAP, and symbol-by-symbol Helstrom followed by classical BP. (b) The same plot as (a) except that the error rates are also displayed in log scale.}
\end{figure}

We construct an explicit quantum circuit for a BPQM-based joint-detection receiver (blueprint shown under Methods, see Fig.~\ref{fig:BPQM_full_circuit}), and prove that it achieves the Helstrom limit for discriminating between the $8$ codewords in our exemplary $n=5$ linear BPSK code with a tree factor graph, as shown in Fig.~\ref{fig:five_bit_code}. 
Hence, it outperforms the best achievable performance by the optimal symbol-by-symbol receiver measurement followed by a MAP decision. 
Based on our analysis of BPQM, we introduce a coherent rotation to be performed after decoding bit $1$ as part of our receiver design, which is not part of Renes' original BPQM scheme.
This might be important for generalizing BPQM beyond the specific example considered here (see Remark~\ref{rem:coherent_rotation}).
We explicitly compute the density matrices of quantum messages that are passed, and evaluate the performance of BPQM for this example code. 
For decoding bit $1$, we also derive an analytical expression for the BPQM success probability. 
The ultimate benchmark for decoding a bit is the performance of the Helstrom measurement that optimally distinguishes the density matrices corresponding to the two values of the bit. 
We show that BPQM is optimal for deciding the value of each of the $5$ bits. 
In Fig.~\ref{fig:bpqm_perr_vs_nbar}, we plot performance curves that show the ``global'' performance of BPQM for the $5$-bit code in terms of block (codeword) error rate for the following strategies: 
\begin{enumerate}
    \item[(a)] collective (optimal) Helstrom measurement on all $n=5$ channel outputs of the received codeword,
    
    \item[(b)] BPQM on all channel outputs of the received codeword,
    
    \item[(c)] symbol-by-symbol (optimal) Helstrom measurement followed by classical (optimal) block-MAP decoding, and
    
    \item[(d)] symbol-by-symbol (optimal) Helstrom measurement followed by classical BP decoding.
\end{enumerate}
For the last two schemes, classical decoding is performed for the BSC, with crossover probability $p = \frac12[1-\sqrt{1-\sigma^2}]$, that is induced by measuring each channel output with the Helstrom measurement to discriminate between $\dket{\pm\theta}$.

As expected, the block error probabilities are in increasing order from (a) through (d). 
The plot shows that BPQM is strictly better than the quantum-optimal symbol-by-symbol detection followed by a block-MAP decision at all values of mean photon number per mode, and that it meets the optimal joint Helstrom measurement on the modulated codeword. 
We confirm this optimality analytically by using the fact that the square root measurement (SRM), also called the pretty good measurement (PGM), is optimal for transmitting binary linear codes on the pure-state channel~\cite{Eldar-it00} (discussed under Methods).
More precisely, we calculate the closed-form expression for the SRM block error probability~\cite{Rengaswamy-arxiv20b}, in terms of the classical code and associated cosets, and the density-matrix based expression for the BPQM block error probability for this example code.
Then, for a range of channel parameters we compute the values from these expressions and confirm that they agree up to even $15$ decimal places.
Therefore, while decomposing the SRM itself into an explicit circuit might be challenging, BPQM provides a circuit that still achieves the optimal block error probability for this code.
This is an important result because, it demonstrates that, if we can construct a BPQM receiver, then it will outperform any known physically realizable receiver for this channel. 
We provide more detailed observations on Fig.~\ref{fig:bpqm_perr_vs_nbar} shortly. 
\subsection{Photon Information Efficiency}
\begin{figure}
\centering


\includegraphics[scale=0.75,keepaspectratio]{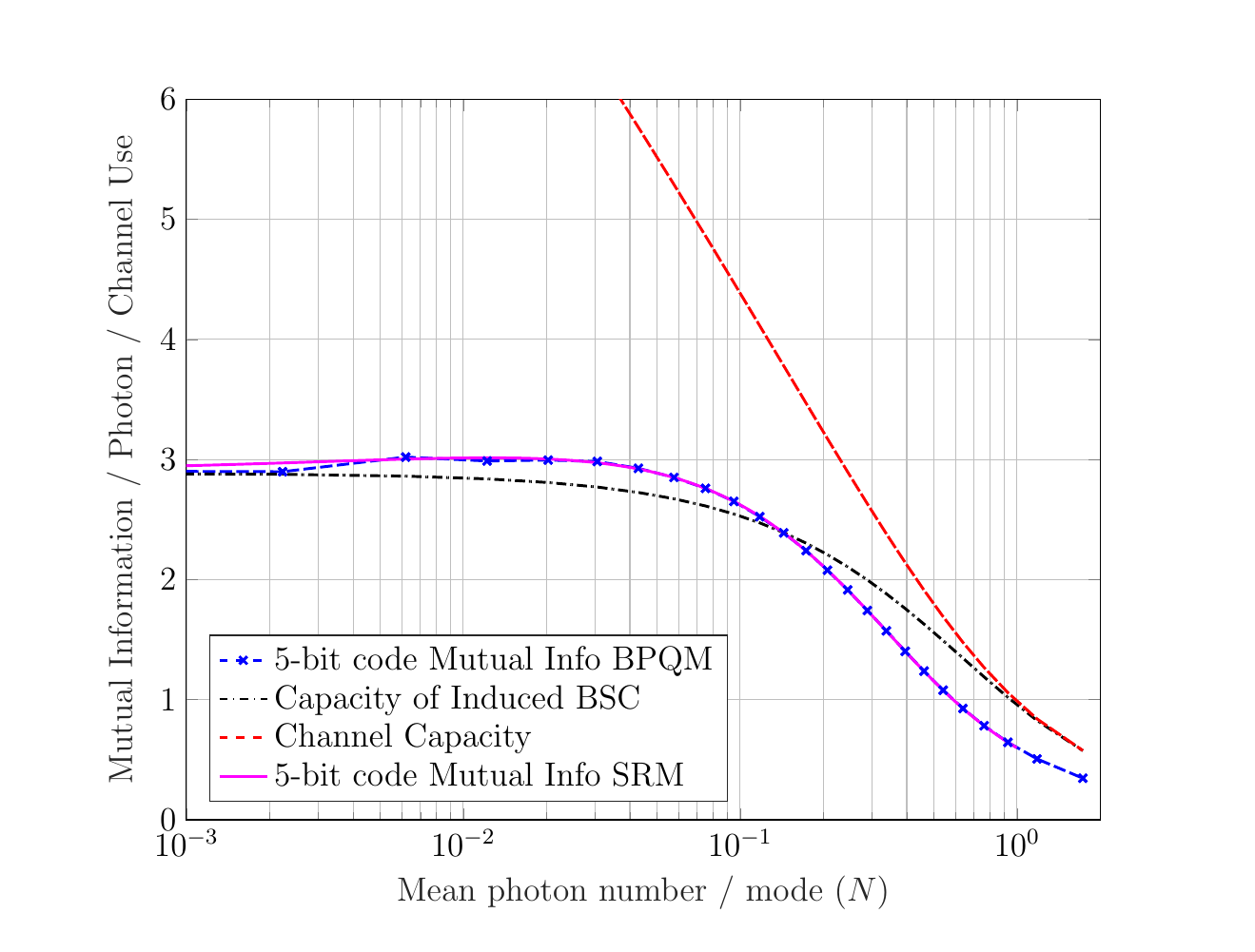}
 
\caption{\label{fig:mi_per_photon} Mutual information per photon per channel use achieved by BPQM and the Square Root Measurement (SRM) on the $5$-bit code over the pure-state channel, along with the Holevo capacity of the channel and the BSC capacity induced by symbol-by-symbol Helstrom measurements at the channel output, all plotted against the mean photon number per mode ($N$). BPQM and SRM on this code produce identical results. The curves indicate that BPQM/SRM provides mutual information gains for certain regimes of $N$, thereby demonstrating superadditive capacity with an explicit code and decoder.}

\end{figure}

Besides the block error rates, we also compare the mutual information-per-photon-per-channel-use, also referred to as the photon information efficiency (PIE)~\cite{Kunz-njp20,Guha-prl11}, for BPQM, with the Holevo capacity of the pure-state channel and the capacity induced by symbol-by-symbol Helstrom measurements.
In order to do this, we consider a composite channel whose input is $k=3$ bits and output is also $k=3$ bits, where the channel consists of encoding into the $5$-bit code, transmitting over the pure-state channel, applying the BPQM receiver and identifying the transmitted codeword (equivalently the $k$-bit message).
Determining the PIE for the BPQM analytically involves calculating closed form expressions for the transition probabilities of the $2^k$-ary channel (where $k=3$ for the considered 5-bit code), which involves cumbersome calculations of the relevant density matrices. 
Instead, we calculate the PIE numerically by performing a Monte Carlo simulation. Fig.~\ref{fig:mi_per_photon} shows a plot of the PIE of the BPQM receiver along with those corresponding to the Holevo capacity and the symbol-by-symbol Helstrom induced BSC capacity.
We also compute the PIE for the square root measurement (SRM). 
The transition probability of decoding a transmitted message $t \in \{0,1\}^k$ as $g \in \{0,1\}^k$ using the SRM is given by~\cite{Rengaswamy-arxiv20b}
\begin{align}
\mathbb{P}\left[ g\, \vert\, t \right] = \frac{ \hat{\sigma}(g \oplus t)^2 }{2^{k}}, \quad \hat{\sigma}(g \oplus t) = \frac{1}{\sqrt{2^k}} \sum_{h \in \mathbb{Z}_2^k} (-1)^{h (g \oplus t)^T} \sigma(h), \quad \sigma(h) = 2^{k/4} \sqrt{\hat{s}(h)},
\end{align}
where $\hat{s}(h)$ is as defined in~\eqref{eq:shat_final} (under Methods, where the block success rate with the SRM given by~\eqref{eq:Psucc_srm} is evaluated).
Therefore, the transition probability only depends on the sum $g \oplus t$ and hence the channel is symmetric.
Using this closed form expression, we compute the mutual information for this $k$-bit channel, normalized by $n=5$ (to obtain mutual information per channel use), then normalized by $N$ to obtain the PIE, and also plot it in Fig.~\ref{fig:mi_per_photon}.
We see that both BPQM and SRM produce identical curves, just as they do in block error rates.
Finally, we observe that there exists a regime of $N$, where this explicit small code along with BPQM or SRM demonstrates superadditive capacity that beats the largest PIE obtained from symbol-by-symbol Helstrom measurements. The PIE with the BPQM (or SRM) receiver is found to be maximized at $N=6.2 \times 10^{-3}$, the maximum PIE being $3.021$, whereas the corresponding PIE attained by symbol-by-symbol Helstrom measurements is $2.862$, the ratio of the two numbers being $1.056$. (Interestingly, this is higher than the PIE ratio of $1.031$ that has been reported for a different $5$-bit code with $16$ codewords and the corresponding SRM in Ref.~\cite{Sasaki-pra98}.) The superadditive PIE hence makes the case stronger for performing the optimal collective measurement at the channel output using the systematic scheme of BPQM.

\begin{figure}
    \centering

	\hspace{-0.75cm}\includegraphics[scale=0.75,keepaspectratio]{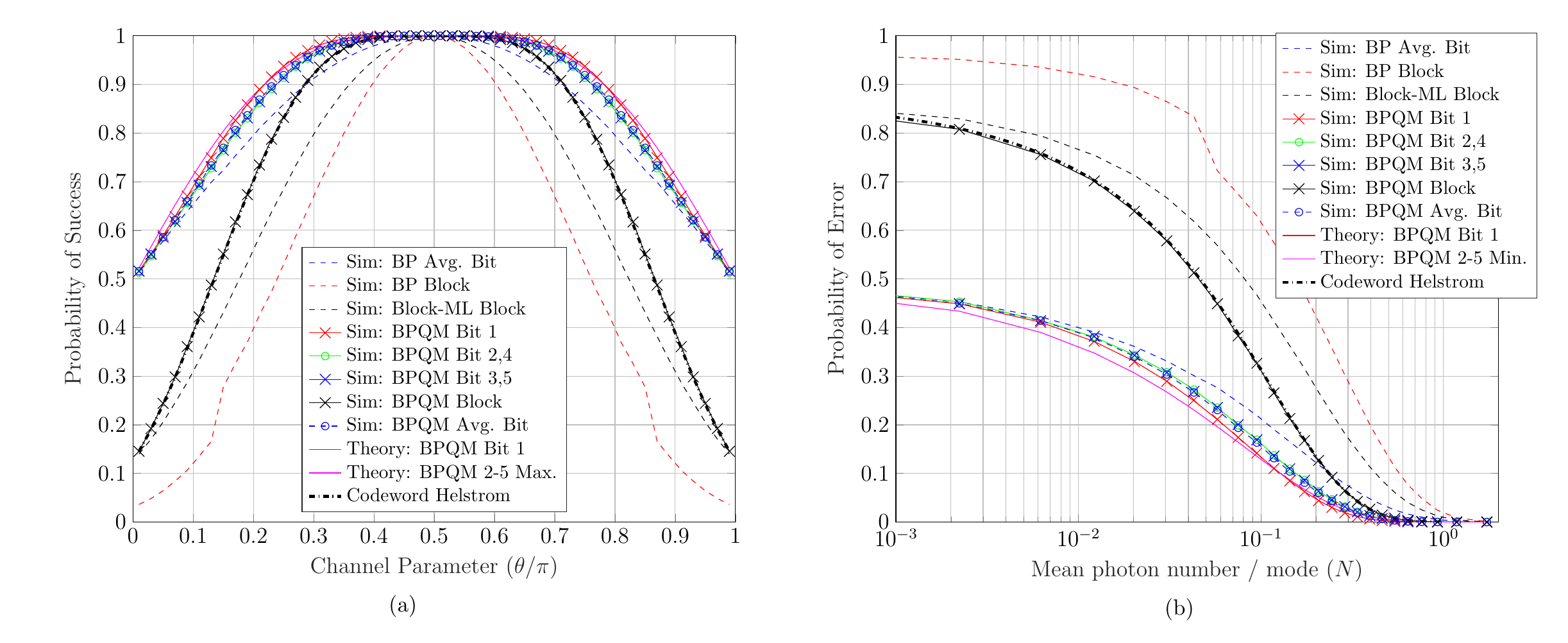}

    \caption{\label{fig:all_bits} The overall performance of BPQM for each bit and for the whole codeword. Each data point in all simulation curves was obtained by averaging over $10^5$ uniformly random codeword transmissions. 
(a) The BPQM success probabilities for decoding each bit and its overall performance for the $5$-bit code, the theoretical BPQM/Helstrom success rate for bit $1$, the initial theoretical prediction of $0.5(1 + \sin\theta^{\circledast})$ for BPQM for bits $2$-$5$, the performance of BP and block-ML when applied to the directly measured channel outputs, and the joint Helstrom limit~\eqref{eq:Psucc_srm}.
(b) The same curves on the left plotted against the mean photon number per mode $N$ ($\cos\theta = e^{-2N}$), where the codeword Helstrom limit was calculated from the YKL limit as in~\cite{Krovi-pra15}. It can be seen that the YKL limit also matches the theoretical calculation from~\eqref{eq:Psucc_srm}.}
\end{figure}


In Fig.~\ref{fig:bpqm_perr_vs_nbar}, we had plotted the block error probabilities as a function of the mean photon number per mode for the different measurement strategies. In Fig.~\ref{fig:all_bits}, we plot both the bit and block error probabilities (and success probabilities, i.e., one minus error probability) for these measurement strategies. For strategy (a), the performance of the collective Helstrom measurement is plotted using the Yuen-Kennedy-Lax (YKL) conditions~\cite{Yuen-it75} as discussed, for example, in~\cite{Krovi-pra15}.
For strategies (c) and (d), classical processing is performed essentially for the BSC induced by measuring each qubit output by the pure-state channel.
The mean photon number per mode, $N$, relates to the pure-state channel parameter $\theta$ as $\cos\theta = e^{-2N}$ (e.g., see~\cite{Guha-isit12} for more information on this quantity).
We make the following observations from these performance curves.
\begin{enumerate}
    \item The block error rates are in increasing order from strategy (a) to (d), as we might expect. Even though classical BP is performed on a tree FG here, it only implements bit-MAP decoding and not block-MAP decoding. This is why it performs worse than block-ML (i.e., block-MAP with uniform prior on codewords) in this case.
    
    \item BPQM performs strictly better than symbol-by-symbol optimal detection followed by classical MAP decoding. This gives a clear demonstration that if one physically constructs a receiver for BPQM, then it will be the best known physically realizable receiver for the pure-state channel. For example, the Dolinar receiver~\cite{Guha-isit12,Dolinar-1973} realizes only strategy (c). One can use our circuits to make such a physical realization. 
    
    \item BPQM performs as well as the quantum optimal collective Helstrom measurement on the outputs of the channel. This lends evidence to the conclusion that by passing quantum messages, BPQM is able to behave like a collective measurement while still making only single-qubit Pauli measurements during the process. However, more careful analysis is required to characterize this in general for, say, the family of codes with tree FGs.
    
    
    \item As a first self-consistency check, we observe that the block-ML curve asymptotes at roughly $0.875$ for low mean photon numbers per mode. This is because, in this regime, the BSC induced by the symbol-by-symbol measurement essentially has a bit-flip rate of $0.5$. Therefore, block-ML computes a posterior that is almost uniform on all codewords, and thus the block success probability is $1/|\MCC| = 1/8 = 0.125$.
    
    \item As another self-consistency check, we note that the BP curve asymptotes at roughly $(1 - 1/32) = 0.9688$ for low mean photon numbers per mode. Since BP performs bit-MAP on this FG, and the induced BSC in this regime flips bits at a rate of almost $0.5$, BP essentially picks each bit uniformly at random, thereby returning a vector that is uniformly at random out of all the possible $2^5 = 32$ vectors of length $5$.
    
    \item The bit error probability plots show that even though BPQM is optimal for bits $x_2$ through $x_5$, it still performs slightly poorly when compared to the performance for $x_1$. This might be attributed to the fact that in the chosen parity-check matrix, bit $x_1$ is involved in both checks whereas the other bits are involved in exactly one of the two checks.
\end{enumerate}

CQ polar codes are known to achieve capacity on CQ channels when paired with a quantum successive cancellation decoder~\cite{Guha-isit12,Wilde-it13*2}. 
It remains to be seen if the same is also true for a BPQM-based decoder. 
Though this is mentioned in~\cite{Renes-njp17}, we think this requires more details in the form of an explicit proof.
The quantum optimality of BPQM shown in this paper for the example $5$-bit code bodes well for BPQM in this regard.
It also remains open as to how BPQM can be generalized to FGs with cycles and also for decoding over general CQ channels. 
We have shown that the coherent rotation we introduced after measuring the first bit plays an important role in BPQM's optimality (see Remark~\ref{rem:coherent_rotation}).
Hence, one needs to refine Renes' definition of the BPQM algorithm in more general settings.
BPQM also has close connections with the recently introduced notion of channel duality~\cite{Renes-arxiv17}. 
The resulting entropic relations could help characterize the performance of a code over a channel using the performance of its dual code over the dual channel.
Since the dual of the pure-state channel is the classical BSC, we believe it may be possible to extend classical techniques for analyzing BP (on BSC), such as density evolution, to analyze BPQM as well.

Leveraging optical realizations of ``cat basis'' quantum logic, i.e., single- and two-qubit quantum gates in the span of coherent states $\dket{\beta}$ and $\dket{-\beta}$~\cite{Ralph-pra03,Gilchrist-joptb04}, our BPQM quantum circuit can be translated into the first fully-structured optical receiver that would attain the quantum limit of minimum-error discrimination of more than two coherent states. 
Since there is a proven classical-quantum performance gap as discussed above, implementing our receiver on an optical quantum processor provides an alternative proposal for a quantum supremacy experiment that is distinct from the conventional proposals based on random circuits~\cite{Arute-nature19}.

\section{Methods}
\label{methods}


In this section, we will analyze the BPQM algorithm on the example $5$-bit code shown in Fig.~\ref{fig:five_bit_code}.
See Supplementary Note 1 for a review of decoding classical linear codes using the belief propagation (BP) algorithm.
It is very useful to interpret the node operations in BP as performing local statistical inference over certain induced channels, and this perspective is explained in Supplementary Note 2.
This interpretation is also extended to classical-quantum channels as first described by Renes in~\cite{Renes-njp17}.
Then Supplementary Note 3 introduces the pure-state channel and describes the BPQM algorithm via its node operations.
The relevant node convolution operations are performed in detail in Supplementary Note 6 for completeness.

\subsection{Decoding Bit $1$}
\label{sec:decode_bit_1}

Let us begin by describing the procedure to decode bit $1$ of the $5$-bit code from Fig.~\ref{fig:five_bit_code}. 
Observe that the codewords belonging to the code are
\begin{align}
\MCC = \{ 00000, 00011, 01100, 01111, 10101, 10110, 11001, 11010 \}.
\end{align}
We assume that all the codewords are equally likely to be transmitted, just as in classical BP. Then the task of decoding the value of the first bit $x_1$ involves distinguishing between the density matrices $\rho_1^{(0)}$ and $\rho_1^{(1)}$, which are uniform mixtures of the states corresponding to the codewords that have $x_1 = 0$ and $x_1 = 1$, respectively, i.e.,
\begin{align}
\rho_1^{(0)} & = \dketbra{\theta}_1 \otimes \frac{1}{4} \bigg[ \dketbra{\theta}_2 \otimes \dketbra{\theta}_3 \otimes \dketbra{\theta}_4 \otimes \dketbra{\theta}_5 + \dketbra{\theta}_2 \otimes \dketbra{\theta}_3 \otimes \dketbra{-\theta}_4 \otimes \dketbra{-\theta}_5 \nonumber \\
  & \hspace*{2cm} + \dketbra{-\theta}_2 \otimes \dketbra{-\theta}_3 \otimes \dketbra{\theta}_4 \otimes \dketbra{\theta}_5 + \dketbra{-\theta}_2 \otimes \dketbra{-\theta}_3 \otimes \dketbra{-\theta}_4 \otimes \dketbra{-\theta}_5 \bigg], \\
\rho_1^{(1)} & = \dketbra{-\theta}_1 \otimes \frac{1}{4} \bigg[ \dketbra{\theta}_2 \otimes \dketbra{-\theta}_3 \otimes \dketbra{\theta}_4 \otimes \dketbra{-\theta}_5 + \dketbra{\theta}_2 \otimes \dketbra{-\theta}_3 \otimes \dketbra{-\theta}_4 \otimes \dketbra{\theta}_5 \nonumber \\
  & \hspace*{3cm} + \dketbra{-\theta}_2 \otimes \dketbra{\theta}_3 \otimes \dketbra{\theta}_4 \otimes \dketbra{-\theta}_5 + \dketbra{-\theta}_2 \otimes \dketbra{\theta}_3 \otimes \dketbra{-\theta}_4 \otimes \dketbra{\theta}_5 \bigg].
\end{align}
These density matrices can be written in terms of the FN channel convolution in (19) of the Supplementary material 
as $\rho_1^{x_1} =\rho_{\pm}= \dketbra{\pm \theta}_1 \otimes [W \boxast W](x_1)_{23} \otimes [W \boxast W](x_1)_{45}$, where we use the notation $\pm \equiv (-1)^{x_1}, \ x_1\in\{0,1\}$.

\begin{figure}[ht]
\begin{center}

\includegraphics[scale=1.1,keepaspectratio]{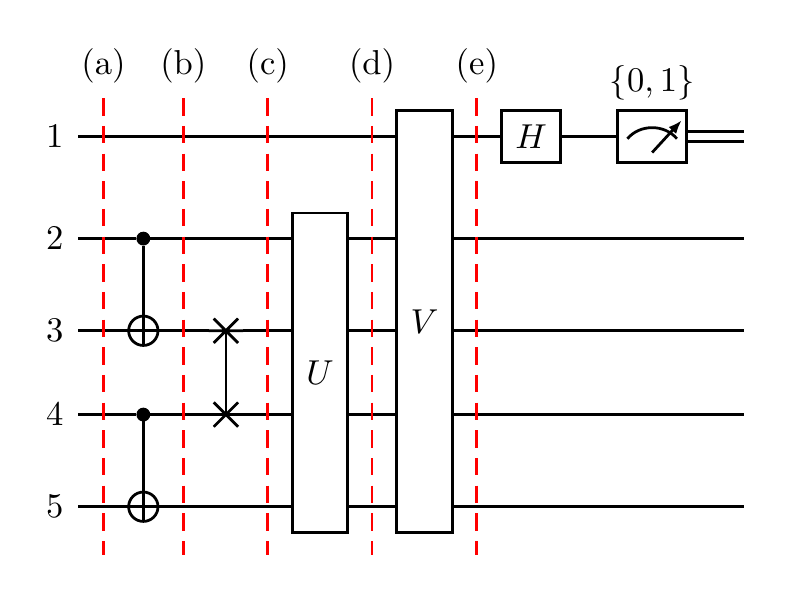}

\caption{\label{fig:BPQM_circuit_bit1}The BPQM circuit to decode bit $1$ of the $5$-bit code in Fig.~\ref{fig:five_bit_code}. All circuits are drawn using the ``Quantikz'' package~\cite{Kay-arxiv18}.}

\vspace*{-0.4cm}

\end{center}
\end{figure}

The BPQM circuit for decoding $x_1$ is shown in Fig.~\ref{fig:BPQM_circuit_bit1} along with the density matrix in each stage of the circuit denoted by (a) through (e).

\begin{itemize}

\item[(a)] $\rho_{\pm,a} = \dketbra{\pm \theta}_1 \otimes [W \boxast W](x_1)_{23} \otimes [W \boxast W](x_1)_{45}$.

\item[(b)] $\rho_{\pm,b} = \dketbra{\pm \theta}_1 \otimes \sum_{j \in \{0,1\}} p_j \dketbra{\pm \theta_j^{\boxast}}_2 \otimes \dketbra{j}_3 \otimes \sum_{k \in \{0,1\}} p_k \dketbra{\pm \theta_k^{\boxast}}_4 \otimes \dketbra{k}_5$.

\item[(c)] $\rho_{\pm,c} = \dketbra{\pm \theta}_1 \otimes \sum_{j,k \in \{0,1\}^2} p_j p_k \dketbra{\pm \theta_j^{\boxast}}_2 \otimes \dketbra{\pm \theta_k^{\boxast}}_3 \otimes \dketbra{j}_4 \otimes \dketbra{k}_5$.

\item[(d)] $\sigma_{\pm} = \sum_{j,k \in \{0,1\}^2} p_j p_k \dketbra{\pm \theta}_1 \otimes \dketbra{\pm \theta_{jk}^{\circledast}}_2 \otimes \dketbra{0}_3 \otimes \dketbra{jk}_{45}$, where the applied unitary operation is $U \coloneqq \sum_{j,k \in \{0,1\}^2} U_{\circledast}(\theta_j^{\boxast}, \theta_k^{\boxast})_{23} \otimes \dketbra{jk}_{45}$ and $\cos\theta_{jk}^{\circledast} \coloneqq \cos\theta_j^{\boxast} \cos\theta_k^{\boxast}$.

\item[(e)] $\Psi_{\pm} = \sum_{j,k \in \{0,1\}^2} p_j p_k \dketbra{\pm \varphi_{jk}^{\circledast}}_1 \otimes \dketbra{0}_2 \otimes \dketbra{0}_3 \otimes \dketbra{jk}_{45}$, where the applied unitary operation is $V \coloneqq \sum_{j,k \in \{0,1\}^2} U_{\circledast}(\theta, \theta_{jk}^{\circledast})_{12} \otimes \dketbra{jk}_{45}$ and $\cos\varphi_{jk}^{\circledast} \coloneqq \cos\theta \cos\theta_{jk}^{\circledast}$.

\end{itemize}
We emphasize that at each stage, the density matrix is the expectation over all pure states obtained there that correspond to transmitted codewords with the first bit taking value $x_1 \in \{ 0,1 \}$.
The operations $U$ and $V$ are effectively two-qubit unitary operations, albeit controlled ones, and this phenomenon extends to any factor graph.
Evidently, BPQM compresses all the quantum information into system $1$ and the problem reduces to distinguishing between {$\Psi_{\pm}^{(1)} = \sum_{j,k \in \{0,1\}^2} p_j p_k \dketbra{\pm \varphi_{jk}^{\circledast}}_1$}, since the other systems are either trivial or completely classical and independent of $x_1$. Finally, system $1$ is measured by projecting onto the Pauli $X$ basis, which we know from the discussion in Supplementary Note 3.2 
after~(31) 
to be the Helstrom measurement to optimally distinguish between the states $\Psi_{\pm}^{(1)}$. 

\begin{figure}
\begin{center}


\includegraphics[scale=0.75,keepaspectratio]{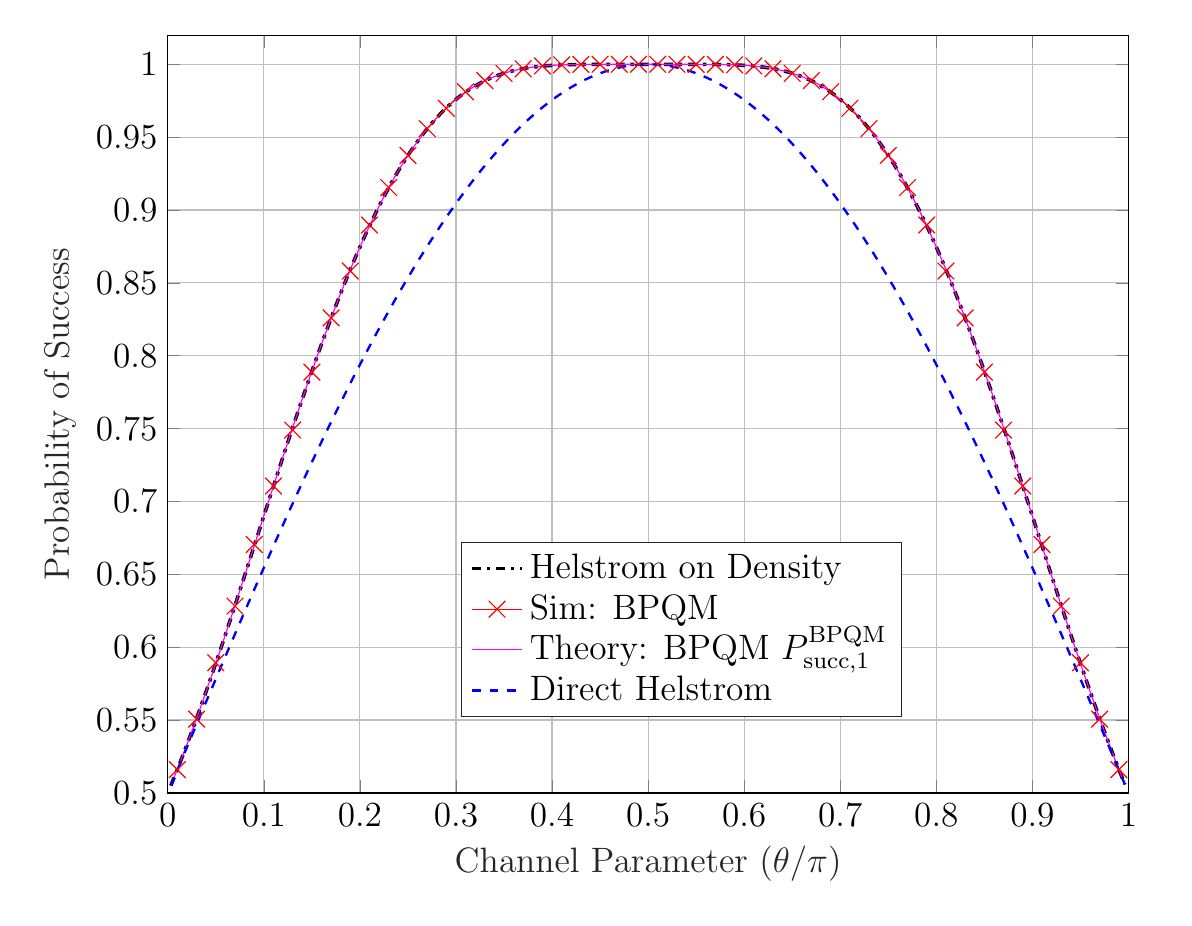}

\caption{\label{fig:bit1_BPQM}The success probabilities for decoding the value of $x_1$ in the $5$-bit code. Here, ``Helstrom on Density'' represents $P_{\text{succ},1}^{\text{Hel}}$ and ``Direct'' represents the success probability when directly implementing the Helstrom measurement at the channel output on system $1$. The curve ``Sim: BPQM'' corresponds to a simulation that averaged each data point over $10^5$ codewords.}

\end{center}
\end{figure}

It is pertinent that the optimal success probability of distinguishing between the density matrices $\rho_1^{(0)}$ and $\rho_1^{(1)}$ using a collective Helstrom measurement is given by
\begin{align}
P_{\text{succ},1}^{\text{Hel}} = \frac{1}{2} + \frac{1}{4} \norm{\rho_1^{(0)} - \rho_1^{(1)}}_1, \ \ \norm{M}_1 \coloneqq \text{Tr}\left( \sqrt{M^{\dagger} M} \right).
\end{align}
The action of BPQM until the final measurement is unitary and the trace norm $\norm{\cdot}_1$ is invariant under unitaries. Thus, BPQM does not lose optimality until the final measurement. Since the final measurement is also optimal for distinguishing the two possible states at that stage (e), BPQM is indeed optimal in decoding the value of $x_1$. Thus, despite not performing a collective measurement, but rather only a single-qubit measurement at the end of a sequence of unitaries motivated by the FG structure and induced channels in classical BP, BPQM is still optimal to determine whether $x_1 = 0$ or $1$. The performance curves plotted in Fig.~\ref{fig:bit1_BPQM} demonstrate this optimality.

\begin{remark}
\normalfont
Observe that in this quantum scenario, $\rho_1^{(x)}$ behave like a ``posterior'' for bit $x_1$.
However, these can be written down even before transmitting over the channel since they do not depend on the output of the channel.
Hence, it is unclear if there is a better notion of a true posterior which we can then show to be ``marginalized'' by BPQM.
\end{remark}




We now analyze the performance of the receiver in decoding bit 1. 
The probability to decode it as $\hat{x}_1 = 0$ is
\begin{align}
\mathbb{P}\left[ \hat{x}_1 = 0\, \vert \, \Psi_{\pm}^{(1)} \right] = \text{Tr}\left[ \Psi_{\pm}^{(1)} \dketbra{+} \right] = \sum_{j,k \in \{0,1\}^2} p_j p_k \left( \frac{1 \pm \sin\varphi_{jk}^{\circledast}}{2} \right).
\end{align}
Therefore, since there are $4$ codewords each that have $x_1 = 0$ and $x_1 = 1$, the prior for bit $x_1$ is $1/2$ and the probability of success for BPQM in decoding the bit $x_1$ is
\begin{align}
P_{\text{succ},1}^{\text{BPQM}} & = \mathbb{P}[x_1 = 0] \cdot \mathbb{P}[\hat{x}_1 = 0 \, | \, x_1 = 0] + \mathbb{P}[x_1 = 1] \cdot \mathbb{P}[\hat{x}_1 = 1 \, | \, x_1 = 1] \\
  & = \frac{1}{2} \left[ \sum_{j,k \in \{0,1\}^2} p_j p_k \left( \frac{1 + \sin\varphi_{jk}^{\circledast}}{2} \right) + \sum_{j,k \in \{0,1\}^2} p_j p_k \left( \frac{1 + \sin\varphi_{jk}^{\circledast}}{2} \right) \right] \\
  & = \frac{p_0^2}{2} \left( 1 + \sin\varphi_{00}^{\circledast} \right) + (1 - p_0^2) \\
  & = 1 - \frac{p_0^2}{2} (1 - \sin\varphi_{00}^{\circledast}),
\end{align}
where we have used the fact that since all channels are identically $W$, we have $\cos\theta_1^{\boxast} = 0$.
We can calculate
\begin{align}
\cos\varphi_{00}^{\circledast} & = \cos\theta \cos\theta_{00}^{\circledast} = \cos\theta \cos^2\theta_0^{\boxast} = \cos\theta \frac{4 \cos^2\theta}{(1 + \cos^2\theta)^2} \\
\Rightarrow \sin\varphi_{00}^{\circledast} & = \sqrt{1 - \frac{16 \cos^6\theta}{(1 + \cos^2\theta)^4}} = \sqrt{1 - \frac{(2p_0 - 1)^3}{p_0^4}} = \frac{\sqrt{p_0^4 - (2p_0 - 1)^3}}{p_0^2}.
\end{align}
Substituting back, we get the BPQM probability of success for bit $x_1$ to be
\begin{align}
P_{\text{succ},1}^{\text{BPQM}} = 1 - \frac{p_0^2 - \sqrt{p_0^4 - (2p_0 - 1)^3}}{2} = P_{\text{succ},1}^{\text{Hel}},
\end{align}
which is the curve plotted as ``Theory: BPQM $P_{\text{succ},1}^{\text{BPQM}}$'' in Fig.~\ref{fig:bit1_BPQM}.

Before measuring system $1$, the state of system $1$ is essentially $\Psi_{\pm}^{(1)} = p_0^2 \dketbra{\pm \varphi_{00}^{\circledast}} + (1 - p_0^2) \dketbra{\pm}$, since $\cos\varphi_{jk}^{\circledast} = 0$ whenever either $j$ or $k$ equals $1$ (or both) and hence $\dketbra{\pm \varphi_{jk}^{\circledast}} = \dketbra{\pm}$.
So, $p_0^2$ is the probability that the system ``confuses'' the decoder, and projection onto the $X$ basis essentially replaces the system with $\dketbra{m_1}$, where $m_1 = (-1)^{\hat{x}_1} \in \{+,-\}$.
The full post-measurement state is given by quantum mechanics to be
\begin{align}
\Phi_{m_1} = \sum_{j,k \in \{0,1\}^2} p_j p_k \frac{\left| \dbraket{m_1}{\pm \varphi_{jk}^{\circledast}} \right|^2}{\text{Tr}\left[ \Psi_{\pm}^{(1)} \dketbra{m_1} \right]} \dketbra{m_1}_1 \otimes \dketbra{00}_{23} \otimes \dketbra{jk}_{45}.
\end{align}
Note that in Fig.~\ref{fig:BPQM_circuit_bit1}, we need to apply a Hadamard after the $Z$-basis measurement in order to ensure that the effective projector is $H \dketbra{\hat{x}_1} H = \dketbra{m_1}$.

Let us denote the overall unitary operation performed in Fig.~\ref{fig:BPQM_circuit_bit1} until stage (e) as $B_1^{\text{BPQM}}$.
As mentioned earlier, the Helstrom measurement to optimally distinguish between $\rho_1^{(0)}$ and $\rho_1^{(1)}$ is given by the POVM $\{ \Pi_1^{\text{Hel}}, \mathbb{I} - \Pi_1^{\text{Hel}} \}$, where 
\begin{align}
\Pi_1^{\text{Hel}} \coloneqq \sum_{i \colon \lambda_i \geq 0} \dketbra{i}, \ \ (\rho_1^{(0)} - \rho_1^{(1)}) \dket{i} = \lambda_i \dket{i}.
\end{align}
BPQM performs the final Helstrom measurement given by the POVM $\{ \tilde{\Pi}_1^{\text{Hel}}, \mathbb{I} - \tilde{\Pi}_1^{\text{Hel}} \}$, where 
\begin{align}
\tilde{\Pi}_1^{\text{Hel}} \coloneqq \sum_{j \colon \lambda_j \geq 0} \dketbra{j} = \dketbra{+}_1 \otimes (I_{16})_{2345}, \qquad \qquad  (\Psi_{+} - \Psi_{-}) \dket{j} & = \lambda_j \dket{j} \\
\Rightarrow \left[ B_1^{\text{BPQM}} \rho_1^{(0)} \left( B_1^{\text{BPQM}} \right)^{\dagger} - B_1^{\text{BPQM}} \rho_1^{(1)} \left( B_1^{\text{BPQM}} \right)^{\dagger} \right] \dket{j} & = \lambda_j \dket{j} \\
\Rightarrow (\rho_1^{(0)} - \rho_1^{(1)}) \left( B_1^{\text{BPQM}} \right)^{\dagger} \dket{j} & = \lambda_j \left( B_1^{\text{BPQM}} \right)^{\dagger} \dket{j}.
\end{align}
Thus, we can express the eigenvectors for $(\rho_1^{(0)} - \rho_1^{(1)})$ as $\dket{i} = \left( B_1^{\text{BPQM}} \right)^{\dagger} \dket{j}$.
This further implies that 
\begin{align}
    \Pi_1^{\text{Hel}} = \sum_{i \colon \lambda_i \geq 0} \dketbra{i} = \left( B_1^{\text{BPQM}} \right)^{\dagger} \left[ \sum_{j \colon \lambda_j \geq 0} \dketbra{j} \right] B_1^{\text{BPQM}} = \left( B_1^{\text{BPQM}} \right)^{\dagger} \left[ \dketbra{+}_1 \otimes (I_{16})_{2345} \right] B_1^{\text{BPQM}}.
\end{align}
Hence, in order to identically apply the Helstrom measurement $\Pi_1^{\text{Hel}}$, BPQM needs to first apply $B_1^{\text{BPQM}}$, then measure the first qubit in the $X$-basis, and finally invert $B_1^{\text{BPQM}}$ on the post-measurement state $\Phi_{m_1}$ above.
Although this is optimal for bit $1$, next we will see that it is beneficial to coherently rotate $\Phi_{m_1}$ before inverting $B_1^{\text{BPQM}}$, which sets up a better state discrimination problem for decoding bit $2$.

\subsection{Decoding Bits $2$ and $3$ (or $4$ and $5$)}
\label{sec:BPQM_bit2}

Next, in order to execute BPQM to decode bit $x_2$, we would ideally hope to change the state $\Phi_{m_1}$ back to the channel outputs.
However, this is impossible after having performed the measurement.
In the original BPQM algorithm~\cite{Renes-njp17}, the procedure to be performed at this stage is ambiguous, so we describe a strategy that treads closely along the path of performing the Helstrom measurement for bit $2$ as well, i.e., optimally distinguishing $\rho_2^{(0)}$ and $\rho_2^{(1)}$ evolved through $\tilde{A}_1^{\text{BPQM}} \coloneqq \left( B_1^{\text{BPQM}} \right)^{\dagger} \left[ \dketbra{m_1}_1 \otimes (I_{16})_{2345} \right] B_1^{\text{BPQM}}$.

In order to be able to run BPQM for bit $x_1$ in reverse to get ``as close'' to the channel outputs as possible, we need to make sure that the state $\Phi_{m_1}$ is modified to be compatible with the (angles used to define the) unitaries $V$ and $U$ in Fig.~\ref{fig:BPQM_circuit_bit1}.
Since we can keep track of the intermediate angles deterministically, we can conditionally rotate subsystem $1$ to be $\dketbra{m_1 \varphi_{00}^{\circledast}}_1$ for $\dketbra{jk}_{45} = \dketbra{00}_{45}$.
Note again that in $\Psi_{\pm}$, when either of $j$ or $k$ is $1$ (or both), $\varphi_{jk}^{\circledast} = \mathrm{\pi}/2$ and hence $\dketbra{\pm \varphi_{jk}^{\circledast}} = \dketbra{\pm}$.
Therefore, if $\hat{x}_1$ is the wrong estimate for $x_1$, then $\dbraket{m_1}{\pm} = 0$ and the superposition in $\Phi_{m_1}$ collapses to a single term with $j = k = 0$.

More precisely, we can implement the unitary operation (see Supplementary Note~5 
for a decomposition of $K_{m_1}$)
\begin{align}
\label{eq:cond_rotation}
M_{m_1} \coloneqq (K_{m_1})_1 \otimes \dketbra{00}_{45} + (I_2)_1 \otimes \left( \dketbra{01}_{45} + \dketbra{10}_{45} + \dketbra{11}_{45} \right),
\end{align}
where $K_+$ and $K_-$ are unitaries chosen to satisfy $K_+ \dket{+} = \dket{\varphi_{00}^{\circledast}}$ and $K_- \dket{-} = \dket{-\varphi_{00}^{\circledast}}$, respectively.
We can easily complete these partially defined unitaries with the conditions $K_+ \dket{-} = \sin\frac{\varphi_{00}^{\circledast}}{2} \dket{0} - \cos\frac{\varphi_{00}^{\circledast}}{2} \dket{1}$ and $K_- \dket{+} = \sin\frac{\varphi_{00}^{\circledast}}{2} \dket{0} + \cos\frac{\varphi_{00}^{\circledast}}{2} \dket{1}$.
Applying $M_{m_1}$ to $\Phi_{m_1}$ we get the desired state (compare to state $\Psi_{\pm}$ in stage (e) of Fig.~\ref{fig:BPQM_circuit_bit1})
\begin{align}
\tilde{\Psi}_{m_1} = \sum_{j,k \in \{0,1\}^2} p_j p_k \frac{\left| \dbraket{m_1}{\pm \varphi_{jk}^{\circledast}} \right|^2}{\text{Tr}\left[ \Psi_{\pm}^{(1)} \dketbra{m_1} \right]} \dketbra{m_1 \varphi_{jk}^{\circledast}}_1 \otimes \dketbra{00}_{23} \otimes \dketbra{jk}_{45}.
\end{align}
Now the BPQM circuit for bit $x_1$, shown in Fig.~\ref{fig:BPQM_circuit_bit1}, can be run in reverse from before the final measurement, i.e., from stage (e) back to stage (a).
Hence, the overall operation on the input state in Fig.~\ref{fig:BPQM_circuit_bit1} is
\begin{align}
\label{eq:A1_BPQM}
 A_1^{\text{BPQM}} \coloneqq \left( B_1^{\text{BPQM}} \right)^{\dagger} M_{m_1} \left[ \dketbra{m_1}_1 \otimes (I_{16})_{2345} \right] B_1^{\text{BPQM}}. 
\end{align}
Then we expect the state to be almost the same as the channel outputs, except that system $1$ will deterministically be in state $\dketbra{m_1 \theta}_1$.
However, a simple calculation shows that this is not completely true since the additional factor $\frac{\left| \dbraket{m_1}{\pm \varphi_{jk}^{\circledast}} \right|^2}{\text{Tr}\left[ \Psi_{\pm}^{(1)} \dketbra{m_1} \right]}$ prevents the density matrix to decompose into a tensor product of two $2$-qubit density matrices at stage (b) of Fig.~\ref{fig:BPQM_circuit_bit1}.
Specifically, when we take $\tilde{\Psi}_{m_1}$ at stage (e) back to stage (b) by inverting the BPQM operations, we arrive at the state
\begin{align}
\tilde{\rho}_{\pm, b}^{(m_1)} & = \dketbra{m_1 \theta}_1 \otimes \sum_{j,k \in \{0,1\}^2} p_j p_k \frac{\left| \dbraket{m_1}{\pm \varphi_{jk}^{\circledast}} \right|^2}{\text{Tr}\left[ \Psi_{\pm}^{(1)} \dketbra{m_1} \right]} \left( \dketbra{m_1 \theta_j^{\boxast}}_{2} \otimes \dketbra{j}_{3} \right) \otimes \left( \dketbra{m_1 \theta_k^{\boxast}}_{4} \otimes \dketbra{k}_{5} \right) \\
  & = 
 \begin{cases}
  \dfrac{1}{P_{\text{succ}, 1}^{\text{BPQM}}} \dketbra{m_1 \theta}_1 \otimes \sum_{j,k \in \{0,1\}^2} p_j p_k \left| \dbraket{m_1}{\pm \varphi_{jk}^{\circledast}} \right|^2 \left( \dketbra{m_1 \theta_j^{\boxast}}_{2} \otimes \dketbra{j}_{3} \right) &  \\
  \hspace{9cm} \otimes \left( \dketbra{m_1 \theta_k^{\boxast}}_{4} \otimes \dketbra{k}_{5} \right)  & \text{if}\ \hat{x}_1 = x_1, \\
  \dketbra{m_1 \theta}_1 \otimes \left( \dketbra{m_1 \theta_0^{\boxast}}_{2} \otimes \dketbra{0}_{3} \right) \otimes \left( \dketbra{m_1 \theta_0^{\boxast}}_{4} \otimes \dketbra{0}_{5} \right)  & \text{if}\ \hat{x}_1 \neq x_1.
 \end{cases} 
\end{align}

\begin{lemma}
Let $C \coloneqq (I_2)_1 \otimes \cnot{2}{3} \otimes \cnot{4}{5}$ and $\dket{\Gamma_{\hat{x}_1}} \coloneqq \cos\frac{\theta_0^{\boxast}}{2} \dket{00} + (-1)^{\hat{x}_1} \sin\frac{\theta_0^{\boxast}}{2} \dket{11}$.
Then
\begin{align}
C \tilde{\rho}_{\pm, b}^{(m_1)} C^{\dagger} & = 
\begin{cases}
\dfrac{1}{P_{\text{succ}, 1}^{\text{BPQM}}} \dketbra{m_1 \theta}_1 \otimes [W \boxast W](\hat{x}_1)_{23} \otimes [W \boxast W](\hat{x}_1)_{45} & \\
 + \dfrac{p_0^2}{P_{\text{succ}, 1}^{\text{BPQM}}} \left[ 0.5 (1 + \sin\varphi_{00}^{\circledast}) - 1 \right] \dketbra{m_1 \theta}_1 \otimes \dketbra{\Gamma_{\hat{x}_1}}_{23} \otimes \dketbra{\Gamma_{\hat{x}_1}}_{45} & \text{if}\ \hat{x}_1 = x_1, \\
 \dketbra{m_1 \theta}_1 \otimes \dketbra{\Gamma_{\hat{x}_1}}_{23} \otimes \dketbra{\Gamma_{\hat{x}_1}}_{45} & \text{if}\ \hat{x}_1 \neq x_1.
\end{cases}
\end{align}
\begin{proof}
We know from the definition of the factor node convolution operation of BPQM that
\begin{align}
C & \left( \dketbra{m_1 \theta}_1 \otimes [W \boxast W](\hat{x}_1)_{23} \otimes [W \boxast W](\hat{x}_1)_{45} \right) C^{\dagger} \nonumber \\
 & = \dketbra{m_1 \theta}_1 \otimes \left( \sum_{j \in \{0,1\}} p_j \dketbra{m_1 \theta_j^{\boxast}}_2 \otimes \dketbra{j}_3 \right) \otimes \left( \sum_{k \in \{0,1\}} p_k \dketbra{m_1 \theta_k^{\boxast}}_4 \otimes \dketbra{k}_5 \right) \\
 & = \rho_{m_1, b}.
\end{align}
This in turn implies that $C \rho_{m_1, b} C^{\dagger} = \dketbra{m_1 \theta}_1 \otimes [W \boxast W](\hat{x}_1)_{23} \otimes [W \boxast W](\hat{x}_1)_{45}$.
Ignoring the first qubit and the constant factor for simplicity, observe that
\begin{align}
\tilde{\rho}_{\pm, b}^{(m_1)} \bigg\vert_{\hat{x}_1 = x_1} & = \sum_{j,k \in \{0,1\}^2} p_j p_k \left( \left| \dbraket{m_1}{\pm \varphi_{jk}^{\circledast}} \right|^2 - 1 + 1 \right) \left( \dketbra{m_1 \theta_j^{\boxast}}_{2} \otimes \dketbra{j}_{3} \right) \otimes \left( \dketbra{m_1 \theta_k^{\boxast}}_{4} \otimes \dketbra{k}_{5} \right)  \\
  & = \rho_{m_1, b} + p_0^2 \left[ 0.5 (1 + \sin\varphi_{00}^{\circledast}) - 1 \right] \left( \dketbra{m_1 \theta_0^{\boxast}}_{2} \otimes \dketbra{0}_{3} \right) \otimes \left( \dketbra{m_1 \theta_0^{\boxast}}_{4} \otimes \dketbra{0}_{5} \right).
\end{align}
We have used the fact that except when $j = k = 0$, assuming $\hat{x}_1 = x_1$, $\dbraket{m_1}{\pm \varphi_{jk}^{\circledast}} = \dbraket{m_1}{m_1 \varphi_{jk}^{\circledast}} = \dbraket{m_1}{m_1} = 1$.
Finally, using $\cnot{2}{3} \left( \dket{m_1 \theta_0^{\boxast}}_{2} \otimes \dket{0}_{3} \right) = \dket{\Gamma_{\hat{x}_1}}$, the result follows for both cases $\hat{x}_1 = x_1$ and $\hat{x}_1 \neq x_1$.
\end{proof}
\end{lemma}

Therefore, after reversing the operations of BPQM for bit $x_1$, the $5$-qubit system is in the state
\begin{align}
\tilde{\rho}_{m_1,a} & = P_{\text{succ}, 1}^{\text{BPQM}} \cdot C \tilde{\rho}_{\pm, b}^{(m_1)} \bigg\vert_{\hat{x}_1 = x_1} C^{\dagger} + \left( 1 - P_{\text{succ}, 1}^{\text{BPQM}} \right) \cdot C \tilde{\rho}_{\pm, b}^{(m_1)} \bigg\vert_{\hat{x}_1 \neq x_1} C^{\dagger} \\
  & = \dketbra{m_1 \theta}_1 \otimes [W \boxast W](\hat{x}_1)_{23} \otimes [W \boxast W](\hat{x}_1)_{45} \nonumber \\
  & \qquad + p_0^2 \left[ 0.5 (1 + \sin\varphi_{00}^{\circledast}) - 1 \right] \dketbra{m_1 \theta}_1 \otimes \dketbra{\Gamma_{\hat{x}_1}}_{23} \otimes \dketbra{\Gamma_{\hat{x}_1}}_{45} \nonumber \\
  & \qquad + \left( 1 - P_{\text{succ}, 1}^{\text{BPQM}} \right) \cdot \dketbra{m_1 \theta}_1 \otimes \dketbra{\Gamma_{\hat{x}_1}}_{23} \otimes \dketbra{\Gamma_{\hat{x}_1}}_{45} \\
  & = \dketbra{m_1 \theta}_1 \otimes [W \boxast W](\hat{x}_1)_{23} \otimes [W \boxast W](\hat{x}_1)_{45},
\end{align}
since $P_{\text{succ}, 1}^{\text{BPQM}} = p_0^2 \cdot 0.5 (1 + \sin\varphi_{00}^{\circledast}) + (1 - p_0^2)$.

\begin{figure}
\begin{center}

\includegraphics[scale=1.1,keepaspectratio]{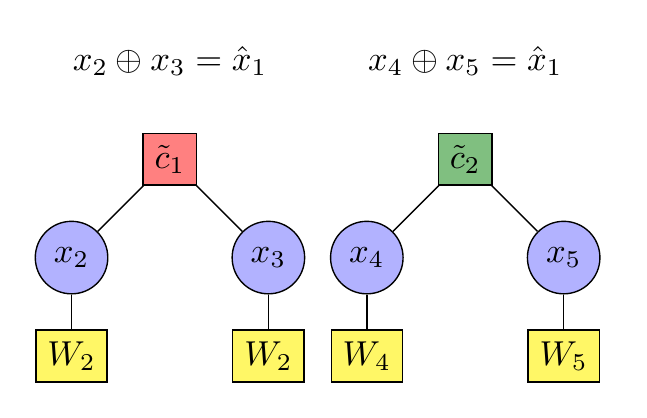}

\caption{\label{fig:fg_post_x1}The reduced factor graph after estimating bit $1$ to be $\hat{x}_1$.}

\end{center}
\end{figure}

At this point, we have decoded $\hat{x}_1 = 0$ if $m_1 = +$ and $\hat{x}_1 = 1$ if $m_1 = -$.
We can absorb the value of $\hat{x}_1$ in the FG by updating the parity checks $c_1$ and $c_2$ to impose $x_2 \oplus x_3 = \hat{x}_1$ and $x_4 \oplus x_5 = \hat{x}_1$, respectively.
Now we have two disjoint FGs as shown in Fig.~\ref{fig:fg_post_x1}.
It suffices to decode $x_2$ and $x_4$ since $\hat{x}_3 = \hat{x}_2 \oplus \hat{x}_1$ and $\hat{x}_5 = \hat{x}_4 \oplus \hat{x}_1$.
Also, due to symmetry, it suffices to analyze the success probability of decoding $x_2$ (resp. $x_4$) and $x_3$ (resp. $x_5$).
For this reduced FG, we need to split $\tilde{\rho}_{m_1, a}$ into two density matrices corresponding to the hypotheses $x_2 = 0$ and $x_2 = 1$.
If we revisit the density matrices $\rho_1^{(0)}$ and $\rho_1^{(1)}$, we observe that the $5$-qubit system at the channel output is exactly $\frac{1}{2} \rho_1^{(0)} + \frac{1}{2} \rho_1^{(1)}$.
Hence, for $x_2$, we accordingly split $[W \boxast W](\hat{x}_1)_{23}$ in $\tilde{\rho}_{m_1, a}$ and arrive at the two hypotheses states 
\begin{align}
\tilde{\Phi}_{x_2 = \hat{x}_1}(\hat{x}_1) & = \dketbra{m_1 \theta}_2 \otimes \dketbra{\theta}_3 \otimes [W \boxast W](\hat{x}_1)_{45}, \\
\tilde{\Phi}_{x_2 \neq \hat{x}_1}(\hat{x}_1) & = \dketbra{-m_1 \theta}_2 \otimes \dketbra{-\theta}_3 \otimes [W \boxast W](\hat{x}_1)_{45}.
\end{align}
In Supplementary Note~4 
we discuss how these hypotheses must be processed, which makes the next steps of BPQM intuitively clear (see Supplementary Figure~3). 
However, the success probability derived from this analysis for bits 2-5 turns out to be significantly higher than the quantum optimal scheme for each bit at the channel output (see Supplementary Figure~4). 
This indicates that the state discrimination problem for bit $2$ discussed above is more ideal than the actual problem in hand.
Hence, next we analyze the true state discrimination problem for bit $2$ (or $4$) and clarify the observed performance in Supplementary Figure~4. 

\begin{figure}
\begin{center}

\includegraphics[scale=1.1,keepaspectratio]{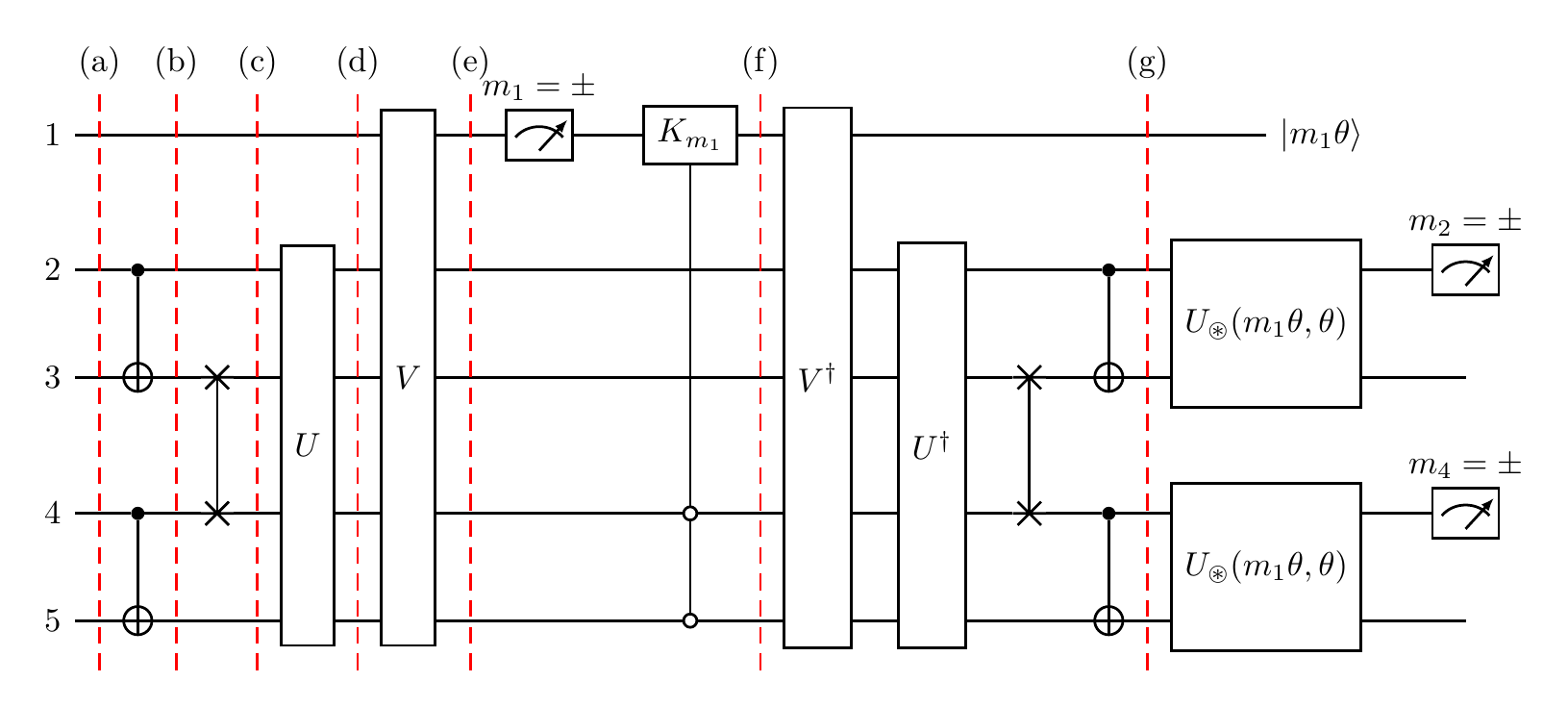}

\caption{\label{fig:BPQM_full_circuit}The full BPQM circuit to decode all bits of the $5$-bit code in Fig.~\ref{fig:five_bit_code}. The decoded values are related to the measurement results as $m_1 = (-1)^{\hat{x}_1}, m_2 = (-1)^{\hat{x}_2}, m_4 = (-1)^{\hat{x}_4}$, and $\hat{x}_3 = \hat{x}_1 \oplus \hat{x}_2, \hat{x}_5 = \hat{x}_1 \oplus \hat{x}_4$. The open-circled controls indicate that $K_{m_1}$ is coherently controlled by the last two qubits being in the state $\dket{00}_{45}$. The solid line before $K_{m_1}$ indicates that the controlled unitary is applied to the post-measurement state. See Fig.~\ref{fig:bpqm_decomposed} for the full decomposition of this circuit.}

\vspace*{-0.4cm}

\end{center}
\end{figure}

\begin{figure}

\centering

\includegraphics[scale=1,keepaspectratio]{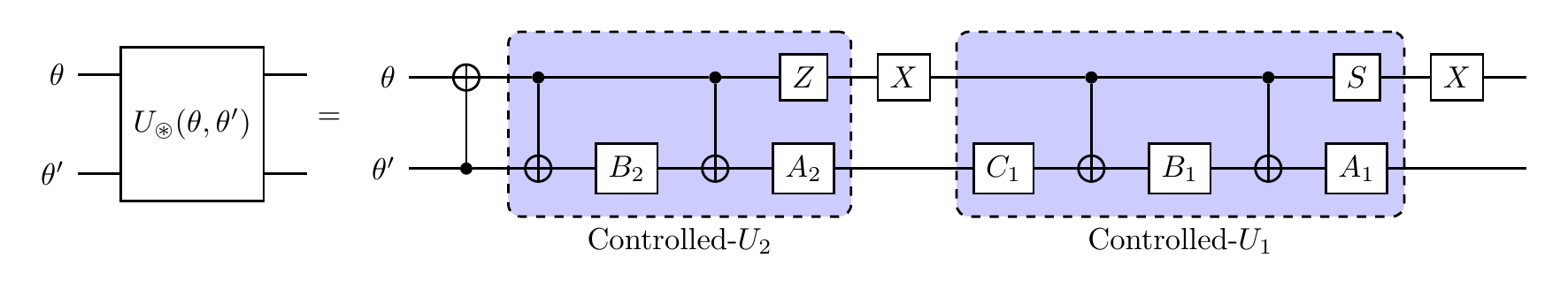}

\caption{\label{fig:vn_unitary_circuit}The circuit decomposition for $U_{\circledast}(\theta,\theta')$, where we set  
$A_1 \coloneqq R_y(\gamma_1/2),  
B_1 \coloneqq R_y(-\gamma_1/2) R_z(-\mathrm{\pi}/2),   
C_1 \coloneqq R_z(\mathrm{\pi}/2), 
A_2 \coloneqq R_z(\mathrm{\pi}) R_y(\gamma_2/2),
B_2 \coloneqq R_y(-\gamma_2/2) R_z(-\mathrm{\pi})$, and $S = \sqrt{Z}$ is the phase gate. See Supplementary Note~5 
for calculations and angles $\gamma_1, \gamma_2$. Note that, for example, $B_2 = R_y(-\gamma_2/2) R_z(-\mathrm{\pi})$ implies that $R_z(-\mathrm{\pi})$ must be applied first, then followed by $R_y(-\gamma_2/2)$.}

\vspace*{-0.2cm}

\end{figure}

\begin{figure}

\centering

\includegraphics[scale=1,keepaspectratio]{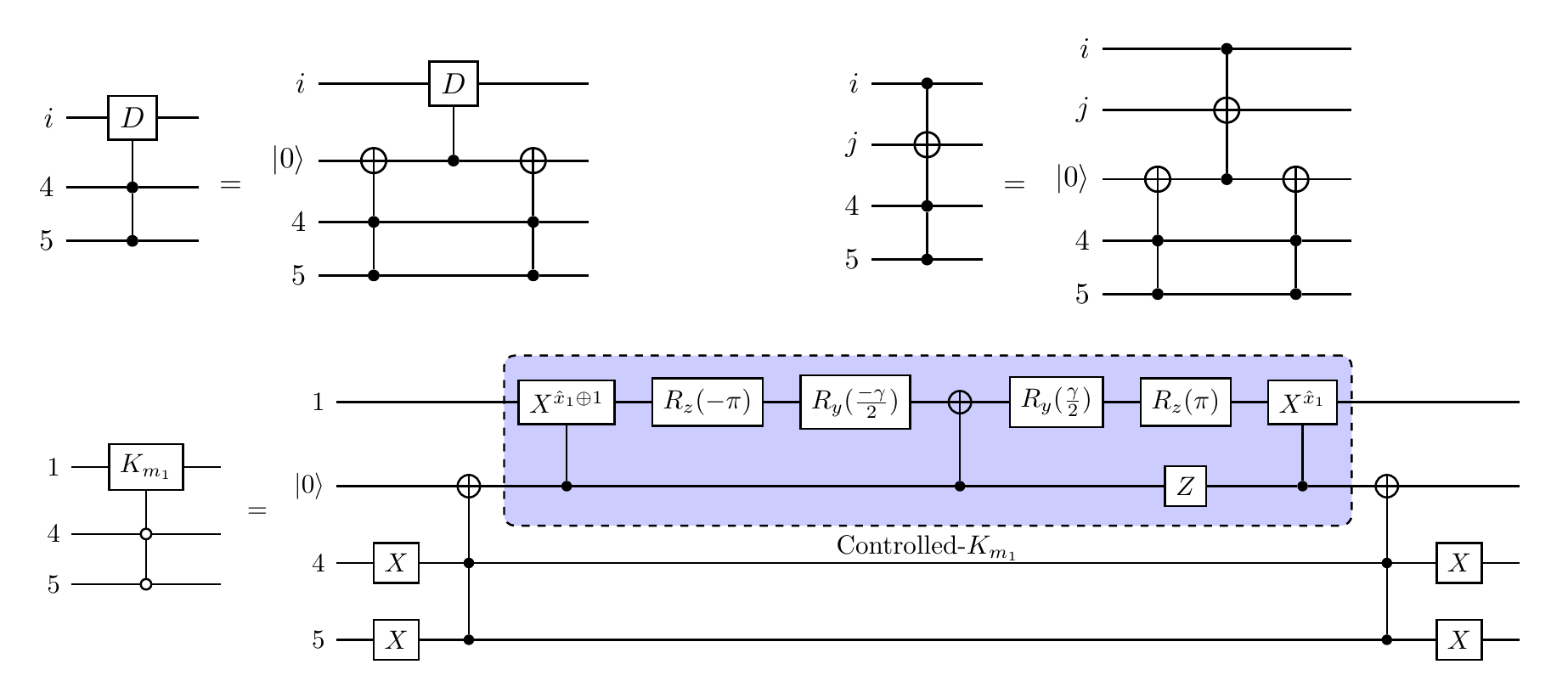}

\caption{\label{fig:controlled_unitary} Decomposition of controlled gates using Toffoli (or CCZ) gates and an ancilla~\cite[Fig. 4.10]{Nielsen-2010}, where $i,j \in \{1,2,3\}$ and $D \in \{A_1,B_1,C_1,A_2,B_2,S\}$ as appropriate. The top two identities can be used to implement each of the doubly-controlled $U_{\circledast}(\theta,\theta')$ appearing in Fig.~\ref{fig:bpqm_decomposed} by applying doubly-controlled versions of the components of $U_{\circledast}(\theta,\theta')$ in Fig.~\ref{fig:vn_unitary_circuit}. Note that $m_1 = (-1)^{\hat{x}_1}$ is the result of estimating $x_1$ to be $\hat{x}_1 \in \{0,1\}$. See Supplementary Note~5 
for the relevant calculations and the angle $\gamma$.}

\vspace{-0.5cm}
    
\end{figure}

\begin{sidewaysfigure}

\centering

\includegraphics[scale=1,keepaspectratio]{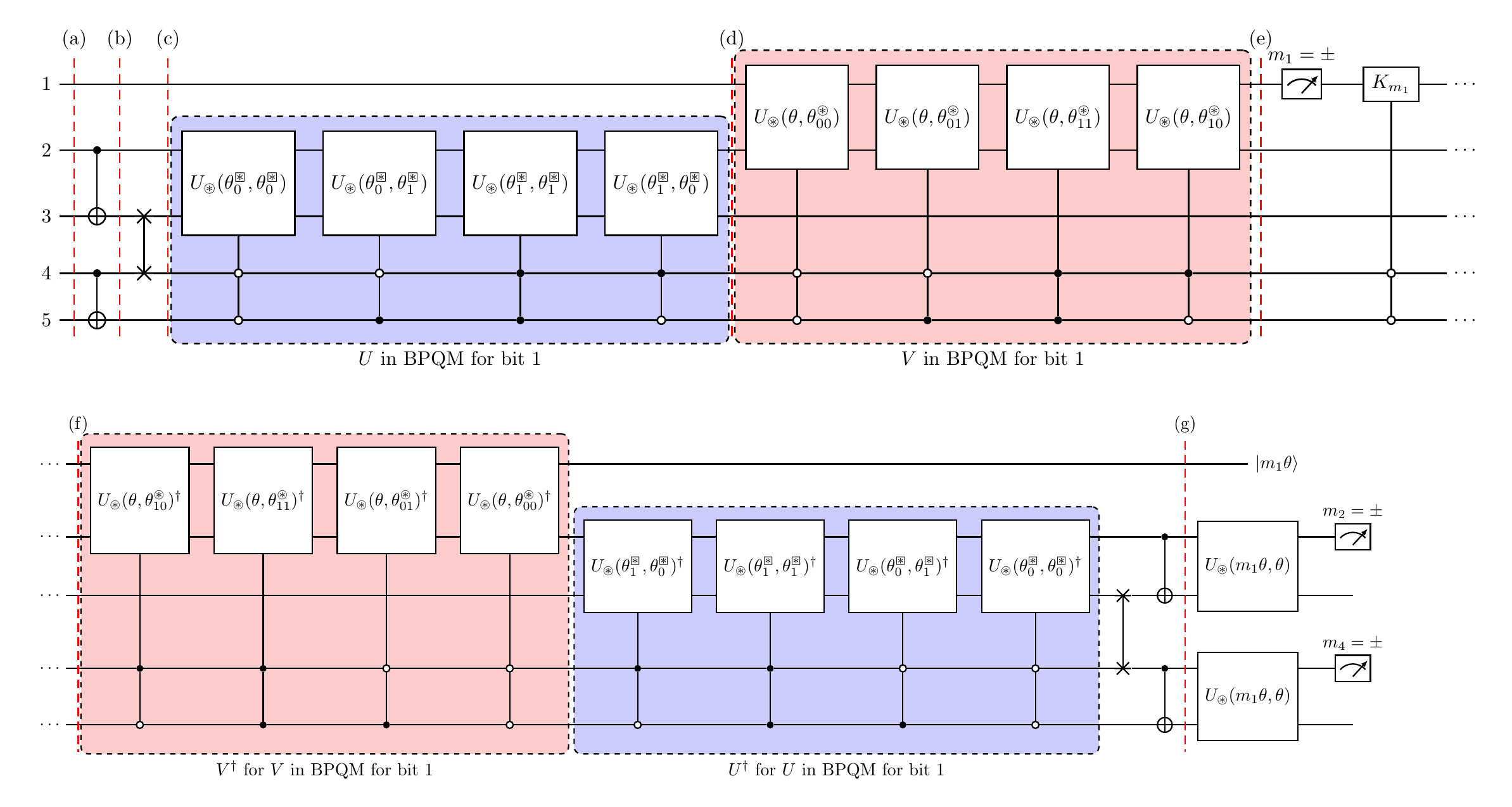}

\caption{\label{fig:bpqm_decomposed} Full decomposition of the BPQM circuit in Fig.~\ref{fig:BPQM_full_circuit}. The variable node unitaries $U_{\circledast}(\theta,\theta')$ are decomposed in Fig.~\ref{fig:vn_unitary_circuit}. The two-qubit-controlled coherent versions of these unitaries as well as the single-qubit rotation $K_{m_1}$, which is a function of the measurement result $m_1$, are decomposed in Fig.~\ref{fig:controlled_unitary}.}

\end{sidewaysfigure}

\subsection{Analysis of BPQM Optimality for Decoding Bit $2$ (or $4$)}
\label{sec:bpqm_bit2_optimal}

At the channel output, it is clear that the optimal strategy to decode bit $2$ is to perform the Helstrom measurement that distinguishes between $\rho_2^{(0)}$ and $\rho_2^{(1)}$.
However, since we performed BPQM operations to decode bit $1$ first, these two density matrices would have evolved through that process.
Therefore, the correct analysis is to derive the resulting states and then subject them to the BPQM strategy for decoding bit $2$ that was discussed above.
For simplicity, we only track the density matrix $\rho_2^{(0)}$ through the different stages in Fig.~\ref{fig:BPQM_full_circuit}.
In Fig.~\ref{fig:bpqm_decomposed} we provide the full expanded BPQM circuit for decoding all bits, and this can be turned into a circuit composed of standard quantum operations by using the decompositions in Figs.~\ref{fig:vn_unitary_circuit} and~\ref{fig:controlled_unitary}.
The corresponding states for $\rho_2^{(1)}$ can easily be ascertained from these.
It will be convenient to express 
\begin{align}
\frac{1}{2} \dketbra{\theta}_2 \otimes \dketbra{\theta}_3 & = [W \boxast W](0)_{23} - \frac{1}{2} \dketbra{-\theta}_2 \otimes \dketbra{-\theta}_3, \\
\frac{1}{2} \dketbra{\theta}_2 \otimes \dketbra{-\theta}_3 & = [W \boxast W](1)_{23} - \frac{1}{2} \dketbra{-\theta}_2 \otimes \dketbra{\theta}_3.
\end{align}
For brevity, we will use the notation $\text{CX}_{ij} \coloneqq \cnot{i}{j}$ and $\text{Swap}_{34} \coloneqq \text{CX}_{34}\, \text{CX}_{43}\, \text{CX}_{34}$.
Then we can write
\begin{align}
\rho_{2,a}^{(0)} & = \frac{1}{2} \dketbra{\theta}_1 \otimes \dketbra{\theta}_2 \otimes \dketbra{\theta}_3 \otimes [W \boxast W](0)_{45} + \frac{1}{2} \dketbra{-\theta}_1 \otimes \dketbra{\theta}_2 \otimes \dketbra{-\theta}_3 \otimes [W \boxast W](1)_{45}, \\
\rho_{2,b}^{(0)} & = \dketbra{\theta}_1 \otimes \left[ \sum_{j = 0,1} p_j \dketbra{\theta_j^{\boxast}}_2 \otimes \dketbra{j}_3 - \frac{1}{2} \text{CX}_{23}\, \dketbra{-\theta,-\theta}_{23}\, \text{CX}_{23} \right] \otimes \left[ \sum_{k = 0,1} p_k \dketbra{\theta_k^{\boxast}}_4 \otimes \dketbra{k}_5 \right] \nonumber \\
  & \quad + \dketbra{-\theta}_1 \otimes \left[ \sum_{j = 0,1} p_j \dketbra{-\theta_j^{\boxast}}_2 \otimes \dketbra{j}_3 - \frac{1}{2} \text{CX}_{23}\, \dketbra{-\theta,\theta}_{23}\, \text{CX}_{23} \right] \nonumber \\
  & \hspace{9cm} \otimes \left[ \sum_{k = 0,1} p_k \dketbra{-\theta_k^{\boxast}}_4 \otimes \dketbra{k}_5 \right], \\
\rho_{2,c}^{(0)} & = \dketbra{\theta}_1 \otimes \sum_{j,k \in \{0,1\}^2} p_j p_k \dketbra{\theta_j^{\boxast}}_2 \otimes \dketbra{\theta_k^{\boxast}}_3 \otimes \dketbra{j}_4 \otimes \dketbra{k}_5 \nonumber \\
  & \quad - \frac{1}{2} \dketbra{\theta}_1 \otimes \text{Swap}_{34}\, \left[ \text{CX}_{23} \dketbra{-\theta,-\theta}_{23} \text{CX}_{23} \otimes  \sum_{k = 0,1} p_k \dketbra{\theta_k^{\boxast}}_4 \otimes \dketbra{k}_5 \right] \text{Swap}_{34} \nonumber \\
  & \quad + \dketbra{-\theta}_1 \otimes \sum_{j,k \in \{0,1\}^2} p_j p_k \dketbra{-\theta_j^{\boxast}}_2 \otimes \dketbra{-\theta_k^{\boxast}}_3 \otimes \dketbra{j}_4 \otimes \dketbra{k}_5 \nonumber \\
  & \quad - \frac{1}{2} \dketbra{-\theta}_1 \otimes \text{Swap}_{34}\, \left[ \text{CX}_{23} \dketbra{-\theta,\theta}_{23} \text{CX}_{23} \otimes  \sum_{k = 0,1} p_k \dketbra{-\theta_k^{\boxast}}_4 \otimes \dketbra{k}_5 \right] \text{Swap}_{34}, \\
\rho_{2,e}^{(0)} & = \sum_{j,k \in \{0,1\}^2} p_j p_k \dketbra{ \varphi_{jk}^{\circledast}}_1 \otimes \dketbra{0}_2 \otimes \dketbra{0}_3 \otimes \dketbra{jk}_{45} \nonumber \\
  & \quad + \sum_{j,k \in \{0,1\}^2} p_j p_k \dketbra{- \varphi_{jk}^{\circledast}}_1 \otimes \dketbra{0}_2 \otimes \dketbra{0}_3 \otimes \dketbra{jk}_{45} \nonumber \\
  & \quad - \frac{1}{2} V U \Biggr\{ \dketbra{\theta}_1 \otimes \text{Swap}_{34}\, \left[ \text{CX}_{23} \dketbra{-\theta,-\theta}_{23} \text{CX}_{23} \otimes  \sum_{k = 0,1} p_k \dketbra{\theta_k^{\boxast}}_4 \otimes \dketbra{k}_5 \right] \text{Swap}_{34} \nonumber \\
  & \quad + \dketbra{-\theta}_1 \otimes \text{Swap}_{34}\, \left[ \text{CX}_{23} \dketbra{-\theta,\theta}_{23} \text{CX}_{23} \otimes  \sum_{k = 0,1} p_k \dketbra{-\theta_k^{\boxast}}_4 \otimes \dketbra{k}_5 \right] \text{Swap}_{34} \Biggr\} U^{\dagger} V^{\dagger}.
\end{align}
Next we make an $X$-basis measurement on the first qubit, and for convenience we assume that the measurement result is $m_1 = +$.
The analysis for $m_1 = -$ is very similar and follows by symmetry.
We verified numerically that $\text{Tr}\left[ \dketbra{+}_1 \cdot \rho_{2,e}^{(0)} \right] = 0.5$, which we might intuitively expect since $\rho_{2,e}^{(0)}$ is the density matrix for $x_2 = 0$ and $x_2$ is independent from $x_1$.
Since $m_1 = +$, we follow the measurement with the conditional rotation $M_+$ in~\eqref{eq:cond_rotation} to obtain
\begin{align}
\Phi_{2,m_1 = +}^{(0)} & = \frac{1}{0.5} \biggr[ \sum_{j,k \in \{0,1\}^2} p_j p_k \lvert \dbraket{+}{\varphi_{jk}^{\circledast}} \rvert^2 \dketbra{ \varphi_{jk}^{\circledast}}_1 \otimes \dketbra{0}_2 \otimes \dketbra{0}_3 \otimes \dketbra{jk}_{45} \nonumber \\
  & \qquad \qquad + \frac{p_0^2 (1 - \sin\varphi_{00}^{\circledast})}{2} \dketbra{\varphi_{00}^{\circledast}} \otimes \dketbra{0}_{2} \otimes \dketbra{0}_3 \otimes \dketbra{00}_{45} \nonumber \\
  & \qquad \qquad - \frac{1}{2} M_+ \dketbra{+}_1 VU \Lambda_2^{(0)} U^{\dagger} V^{\dagger} \dketbra{+}_1 M_+^{\dagger} \biggr], \\
\Lambda_2^{(0)} & \coloneqq \dketbra{\theta}_1 \otimes \text{Swap}_{34}\, \left[ \text{CX}_{23} \dketbra{-\theta,-\theta}_{23} \text{CX}_{23} \otimes  \sum_{k = 0,1} p_k \dketbra{\theta_k^{\boxast}}_4 \otimes \dketbra{k}_5 \right] \text{Swap}_{34} \nonumber \\
  & \quad + \dketbra{-\theta}_1 \otimes \text{Swap}_{34}\, \left[ \text{CX}_{23} \dketbra{-\theta,\theta}_{23} \text{CX}_{23} \otimes  \sum_{k = 0,1} p_k \dketbra{-\theta_k^{\boxast}}_4 \otimes \dketbra{k}_5 \right] \text{Swap}_{34}.
\end{align}
This is the state at stage (f) in Fig.~\ref{fig:BPQM_full_circuit}.
Hence, for $x_2 = 0$, the density matrix we have when $\hat{x}_1 = 0$ and we reverse the BPQM operations on $\Phi_{2,m_1 = +}^{(0)}$ is
\begin{align}
\tilde{\rho}_{2,m_1 = +}^{(0)} & = \frac{1}{0.5} \biggr[ \dketbra{\theta}_1 \otimes [W \boxast W](0)_{23} \otimes [W \boxast W](0)_{45} \nonumber \\
  & \qquad \qquad - \frac{1}{2} \text{CX}_{23}\, \text{CX}_{45}\, \text{Swap}_{34} U^{\dagger} V^{\dagger} M_+ \dketbra{+}_1 V U \Lambda_2^{(0)} U^{\dagger} V^{\dagger} \dketbra{+}_1 M_+^{\dagger} V U \text{Swap}_{34}\, \text{CX}_{45}\, \text{CX}_{23} \biggr].
\end{align}
This is the state at stage (g) in Fig.~\ref{fig:BPQM_full_circuit}.
So, this is the actual density matrix that BPQM encounters for $x_2 = 0$ after having estimated $\hat{x}_1 = 0$.
When compared with the earlier analysis, we observe numerically that this is close to $\tilde{\Phi}_{x_2 = \hat{x}_1}(\hat{x}_1)$ but is not exactly the same.
For example, when $\theta = 0.1\mathrm{\pi}$ we find that $\norm{\tilde{\rho}_{2,m_1 = +}^{(0)} - \tilde{\Phi}_{x_2 = 0}(0)}_{\text{Fro}} = 0.0542$, where ``Fro'' denotes the Frobenius norm, and only two of the distinct entries differ (slightly).
Similarly, 
\begin{align}
\tilde{\rho}_{2,m_1 = +}^{(1)} & = \frac{1}{0.5} \biggr[ \dketbra{\theta}_1 \otimes [W \boxast W](0)_{23} \otimes [W \boxast W](0)_{45} \nonumber \\
  & \qquad \qquad - \frac{1}{2} \text{CX}_{23}\, \text{CX}_{45}\, \text{Swap}_{34} U^{\dagger} V^{\dagger} M_+ \dketbra{+}_1 V U \Lambda_2^{(1)} U^{\dagger} V^{\dagger} \dketbra{+}_1 M_+^{\dagger} V U \text{Swap}_{34}\, \text{CX}_{45}\, \text{CX}_{23} \biggr], \\
\Lambda_2^{(1)} & \coloneqq \dketbra{\theta}_1 \otimes \text{Swap}_{34}\, \left[ \text{CX}_{23} \dketbra{\theta,\theta}_{23} \text{CX}_{23} \otimes  \sum_{k = 0,1} p_k \dketbra{\theta_k^{\boxast}}_4 \otimes \dketbra{k}_5 \right] \text{Swap}_{34} \nonumber \\
  & \quad + \dketbra{-\theta}_1 \otimes \text{Swap}_{34}\, \left[ \text{CX}_{23} \dketbra{\theta,-\theta}_{23} \text{CX}_{23} \otimes  \sum_{k = 0,1} p_k \dketbra{-\theta_k^{\boxast}}_4 \otimes \dketbra{k}_5 \right] \text{Swap}_{34}.
\end{align}
However, most importantly, we observe that
$\frac{1}{2} \tilde{\rho}_{2,m_1=+}^{(0)} + \frac{1}{2} \tilde{\rho}_{2,m_1=+}^{(1)} = \frac{1}{2} \tilde{\Phi}_{x_2 = 0}(0) + \frac{1}{2} \tilde{\Phi}_{x_2 = 1}(0)$.
This explains that while the full density matrix $\tilde{\rho}_{m_1,a}$ was correct, we had split it incorrectly to arrive at the two hypotheses $\tilde{\Phi}_{x_2 = \hat{x}_1}(\hat{x}_1)$ and $\tilde{\Phi}_{x_2 \neq \hat{x}_1}(\hat{x}_1)$.
Now, the Helstrom measurement that optimally distinguishes between $\tilde{\rho}_{2,m_1 = +}^{(0)}$ and $\tilde{\rho}_{2,m_1 = +}^{(1)}$ only depends on
\begin{align}
\tilde{\rho}_{2,m_1 = +}^{(0)} - \tilde{\rho}_{2,m_1 = +}^{(1)} & = A \left[ \Lambda_2^{(1)} - \Lambda_2^{(0)} \right] A^{\dagger}, \ \ 
A \coloneqq \text{CX}_{23}\, \text{CX}_{45}\, \text{Swap}_{34} U^{\dagger} V^{\dagger} M_+ \dketbra{+}_1 V U.
\end{align}
By symmetry of $m_1 = +$ and $m_1 = -$, the optimal success probability to decide bit $2$ is given by
\begin{align}
P_{\text{succ},2}^{\text{Hel}} & = \frac{1}{2} + \frac{1}{4} \norm{\tilde{\rho}_{2,m_1 = +}^{(0)} - \tilde{\rho}_{2,m_1 = +}^{(1)}}_1 = \frac{1}{2} + \frac{1}{4} \norm{A \left[ \Lambda_2^{(1)} - \Lambda_2^{(0)} \right] A^{\dagger}}_1 = \frac{1}{2} + \frac{1}{4} \norm{L \left( \rho_{2}^{(0)} - \rho_{2}^{(1)} \right) L^{\dagger}}_1, \\
L & \coloneqq \text{CX}_{23}\, \text{CX}_{45}\, \text{Swap}_{34} U^{\dagger} V^{\dagger} M_+ \frac{\dketbra{+}_1}{\sqrt{0.5}} V U \text{Swap}_{34}\, \text{CX}_{45}\, \text{CX}_{23}.
\end{align}

Since $L$ is not unitary, we cannot directly apply the unitary invariance of the trace norm to conclude that there is no degradation in performance when compared to optimally distinguishing $\rho_{2}^{(0)}$ and $\rho_{2}^{(1)}$ at the channel output.
However, we observe numerically (even up to $12$ significant digits) that the operations in $L$ indeed ensure that $\norm{L \left( \rho_{2}^{(0)} - \rho_{2}^{(1)} \right) L^{\dagger}}_1 = \norm{\rho_{2}^{(0)} - \rho_{2}^{(1)}}_1$.
Moreover, we also observe that the BPQM operations for bit $2$ given in Supplementary Figure~3 
achieve the same success probability, i.e., using the notation $\pm \equiv (-1)^{x_2}$ we have
\begin{align}
P_{\text{succ}, 2}^{\text{BPQM}} & = \text{Tr}\left[ U_{\circledast}(m_1 \theta,\theta) \tilde{\rho}_{2,m_1}^{(x_2)} U_{\circledast}(m_1 \theta,\theta)^{\dagger} \cdot \dketbra{\pm}_2 \right] \\
  & = \frac{1}{2} + \frac{1}{4} \norm{L \left( \rho_{2}^{(0)} - \rho_{2}^{(1)} \right) L^{\dagger}}_1 \\
  & = \frac{1}{2} + \frac{1}{4} \norm{\rho_{2}^{(0)} - \rho_{2}^{(1)}}_1 \\
  & = P_{\text{succ},2}^{\text{Hel}}.
\end{align}
Finally, the simulation results in Fig.~\ref{fig:all_bits} clearly show that the overall block error rate of BPQM coincides with that of the quantum optimal joint Helstrom limit.
It remains open to rigorously prove all of these observations.



\subsection{Overall Performance of BPQM}
\label{sec:BPQM_perf_all_bits}

We can calculate the probability that the full codeword $\vecnot{x}$ is decoded correctly as
\begin{align}
P_{\text{succ}}^{\text{BPQM}} = \mathbb{P}[\vecnot{\hat{x}} = \vecnot{x}] & = \mathbb{P}[\hat{x}_1 = x_1] \cdot \mathbb{P}[\hat{x}_2 = x_2 \, | \, \hat{x}_1 = x_1] \cdot \mathbb{P}[\hat{x}_4 = x_4 \, | \, \hat{x}_1 = x_1, \hat{x}_2 = x_2] \\
  & = P_{\text{succ},1}^{\text{BPQM}} \cdot P_{\text{succ},2}^{\text{BPQM}} \bigg\vert_{\hat{x}_1 = x_1 = 0} \cdot P_{\text{succ},4}^{\text{BPQM}} \bigg\vert_{\substack{\hat{x}_1 = x_1 = 0\\ \hat{x}_2 = x_2 = 0}}\label{overallPsucc}.
\end{align}
The first term in (\ref{overallPsucc}) is clearly $P_{\text{succ},1}^{\text{BPQM}} = P_{\text{succ},1}^{\text{Hel}}$.
The second term, however, is different from $P_{\text{succ},2}^{\text{BPQM}} = P_{\text{succ},2}^{\text{Hel}}$ because of the conditioning on $x_1$ being estimated correctly, whereas in the above analysis we had implicitly averaged over $\hat{x}_1 = x_1$ and $\hat{x}_1 \neq x_1$.
Nevertheless, we can use a similar strategy as above to derive an expression for the second term.
Here, we want to condition on $x_1$ being estimated correctly, i.e., $\hat{x}_1 = x_1$, and derive the hypothesis states for $x_2$ under this scenario.
Similarly, for the third term, the additional conditioning on $\hat{x}_2 = x_2$ makes it not equal to the second term, although $x_2$ and $x_4$ are placed symmetrically in the factor graph of the code.
But it still holds that $P_{\text{succ},2}^{\text{BPQM}} \bigg\vert_{\hat{x}_1 = x_1 = 0} = P_{\text{succ},4}^{\text{BPQM}} \bigg\vert_{\hat{x}_1 = x_1 = 0}$.
We perform these two analyses next and then combine them to calculate the full block success probability of BPQM.


We will first analyze the decoding of bit $2$ conditioned on bit $1$. 
Let $(\rho_2^{(00)}, \rho_2^{(01)}), (\rho_2^{(10)}, \rho_2^{(11)})$ be two pairs of hypothesis states for $x_2$, at the channel output, where the first pair is conditioned on $x_1 = 0$ and the second on $x_1 = 1$, and this information is known to the receiver.
It is clear, for example, that $\rho_{2}^{(0x_2)} = \dketbra{\theta}_1 \otimes \dketbra{(-1)^{x_2} \theta}_2 \otimes \dketbra{(-1)^{x_2} \theta}_3 \otimes [W \boxast W](0)_{45}$.
After similar calculations as before, we finally obtain
\begin{align}
\tilde{\sigma}_{2,m_1 = +}^{(00)} & = \frac{1}{ P_{\text{succ},1}^{\text{Hel}} } \text{CX}_{23}\, \text{CX}_{45}\, \text{Swap}_{34} U^{\dagger} V^{\dagger} \biggr[ 2 \sum_{j,k \in \{0,1\}^2} p_j p_k \lvert \dbraket{+}{\varphi_{jk}^{\circledast}} \rvert^2 \dketbra{ \varphi_{jk}^{\circledast}}_1 \otimes \dketbra{0}_2 \otimes \dketbra{0}_3 \otimes \dketbra{jk}_{45} \nonumber \\
  & \hspace{3.5cm} -  M_+ \dketbra{+}_1 V U \tilde{\Lambda}_2^{(0)} U^{\dagger} V^{\dagger} \dketbra{+}_1 M_+^{\dagger} \biggr] V U \text{Swap}_{34}\, \text{CX}_{45}\, \text{CX}_{23}, \\
\tilde{\Lambda}_2^{(0)} & \coloneqq \dketbra{\theta}_1 \otimes \text{Swap}_{34}\, \left[ \text{CX}_{23} \dketbra{-\theta,-\theta}_{23} \text{CX}_{23} \otimes  \sum_{k = 0,1} p_k \dketbra{\theta_k^{\boxast}}_4 \otimes \dketbra{k}_5 \right] \text{Swap}_{34}.
\end{align}
This is the state at stage (g) in Fig.~\ref{fig:BPQM_full_circuit}.
So, this is the actual density matrix that BPQM encounters for $x_2 = 0$ after having estimated correctly that $\hat{x}_1 = x_1 = 0$ (and reversed the first set of operations).
Similarly,
\begin{align}
\tilde{\sigma}_{2,m_1 = +}^{(01)} & = \frac{1}{ P_{\text{succ},1}^{\text{Hel}} } \text{CX}_{23}\, \text{CX}_{45}\, \text{Swap}_{34} U^{\dagger} V^{\dagger} \biggr[ 2 \sum_{j,k \in \{0,1\}^2} p_j p_k \lvert \dbraket{+}{\varphi_{jk}^{\circledast}} \rvert^2 \dketbra{ \varphi_{jk}^{\circledast}}_1 \otimes \dketbra{0}_2 \otimes \dketbra{0}_3 \otimes \dketbra{jk}_{45} \nonumber \\
  & \hspace{3.5cm} -  M_+ \dketbra{+}_1 V U \tilde{\Lambda}_2^{(1)} U^{\dagger} V^{\dagger} \dketbra{+}_1 M_+^{\dagger} \biggr] V U \text{Swap}_{34}\, \text{CX}_{45}\, \text{CX}_{23}, \\
\tilde{\Lambda}_2^{(1)} & \coloneqq \dketbra{\theta}_1 \otimes \text{Swap}_{34}\, \left[ \text{CX}_{23} \dketbra{\theta,\theta}_{23} \text{CX}_{23} \otimes  \sum_{k = 0,1} p_k \dketbra{\theta_k^{\boxast}}_4 \otimes \dketbra{k}_5 \right] \text{Swap}_{34}.
\end{align}
The Helstrom measurement that optimally distinguishes between $\tilde{\sigma}_{2,m_1 = +}^{(00)}$ and $\tilde{\sigma}_{2,m_1 = +}^{(01)}$ achieves the success probability 
\begin{align}
P_{\text{succ},2}^{\text{Hel}} \bigg\vert_{\hat{x}_1 = x_1 = 0} & = \frac{1}{2} + \frac{1}{4} \norm{ \tilde{\sigma}_{2,m_1 = +}^{(00)} - \tilde{\sigma}_{2,m_1 = +}^{(01)} }_1 \\
 & = \frac{1}{2} + \frac{1}{4 P_{\text{succ},1}^{\text{Hel}}} \norm{A \left[ \tilde{\Lambda}_2^{(1)} - \tilde{\Lambda}_2^{(0)} \right] A^{\dagger}}_1. 
%
\end{align}
We verified numerically that the final processing of BPQM, after (g) in Fig.~\ref{fig:BPQM_full_circuit}, also achieves the same success probability, i.e., 
\begin{align}
P_{\text{succ},2}^{\text{BPQM}} \bigg\vert_{\hat{x}_1 = x_1 = 0} & = \text{Tr}\left[ U_{\circledast}(\theta,\theta) \tilde{\sigma}_{2,m_1=+}^{(0 x_2)} U_{\circledast}(\theta,\theta)^{\dagger} \cdot \dketbra{(-1)^{x_2}}_2 \right] \\
  & = \frac{1}{2} + \frac{1}{4 P_{\text{succ},1}^{\text{Hel}}} \norm{A \left[ \tilde{\Lambda}_2^{(1)} - \tilde{\Lambda}_2^{(0)} \right] A^{\dagger}}_1 \\
  & = P_{\text{succ},2}^{\text{Hel}} \bigg\vert_{\hat{x}_1 = x_1 = 0}.
\end{align}
Using a similar procedure as above, we can verify the analogous result for $x_2$ conditioned on $x_1 = 1$ and $\hat{x}_1 = x_1 = 1$.



Next, we will analyze the decoding of bit $4$ conditioned on bits $1$ and $2$. 
For convenience, let us assume that $x_1 = x_2 = 0$ in the transmitted codeword.
Note that, due to symmetry, this choice will not affect the analysis and the final probability of success for $x_4$ conditioned on correct estimation of $x_1$ and $x_2$ will be independent of this fixed choice.
Then, at the channel output, the candidate states for $x_4$ are given by
\begin{align}
\rho_4^{(00x_4)} = \dketbra{\theta}_1 \otimes \dketbra{\theta}_2 \otimes \dketbra{\theta}_3 \otimes \dketbra{(-1)^{x_4} \theta}_4 \otimes \dketbra{(-1)^{x_4} \theta}_5.
\end{align}
Let $U_1 = V U \text{Swap}_{34}\, \text{CX}_{45}\, \text{CX}_{23}$ and $\rho_{4,1}^{(00x_4)} = U_1 \rho_4^{(00x_4)} U_1^{\dagger}$.
If $p_{1,4}^{(00x_4)} = \tr{ \dketbra{+}_1 \cdot \rho_{4,1}^{(00x_4)} }$ is the probability of measuring $x_1 = 0$, then conditioned on this correct measurement we arrive at the following candidate states after the next set of BPQM operations:
\begin{align}
\rho_{4,2}^{(00x_4)} & = \frac{ U_2 \cdot \rho_{4,1}^{(00x_4)} U_2^{\dagger} }{p_{1,4}^{(00x_4)}}, \ \ 
U_2 \coloneqq U_{\circledast}(\theta,\theta)_{45}\, U_{\circledast}(\theta,\theta)_{23}\, \text{CX}_{23}\, \text{CX}_{45}\, \text{Swap}_{34} U^{\dagger} V^{\dagger} M_+ \dketbra{+}_1.
\end{align}
If $p_{2,4}^{(00x_4)} = \tr{ \dketbra{+}_2 \cdot \rho_{4,2}^{(00x_4)} }$ is the probability of measuring $x_2 = 0$, conditioned on $\hat{x}_1 = x_1$, then conditioned on this correct measurement we arrive at the following candidate states after the $x_2$ measurement:
\begin{align}
\rho_{4,3}^{(00x_4)} & = \frac{ \dketbra{+}_2 \cdot \rho_{4,1}^{(00x_4)} \dketbra{+}_2 }{p_{2,4}^{(00x_4)}}.
\end{align}
Therefore, any measurement that optimally distinguishes between $x_4 = 0$ and $x_4 = 1$ conditioned on $\hat{x}_1 = x_1$ and $\hat{x}_2 = x_2$ must satisfy the same probability of success as the Helstrom measurement on $(\rho_{4,3}^{(000)}, \rho_{4,3}^{(001)})$.
We verified that measuring the $4$th qubit in the $X$ basis on $\rho_{4,3}^{(00x_4)}$ indeed satisfies this and hence BPQM is optimal in estimating $x_4$ conditioned on estimating $x_1$ and $x_2$ correctly, i.e.,
\begin{align}
P_{\text{succ},4}^{\text{BPQM}} \bigg\vert_{\substack{\hat{x}_1 = x_1 = 0\\ \hat{x}_2 = x_2 = 0}} & = \text{Tr}\left[ \rho_{4,3}^{(00x_4)} \cdot \dketbra{(-1)^{x_4}}_4 \right] \\
  & = \frac{1}{2} + \frac{1}{4} \norm{ \rho_{4,3}^{(000)} - \rho_{4,3}^{(001)} }_1 \\
  & = P_{\text{succ},4}^{\text{Hel}} \bigg\vert_{\substack{\hat{x}_1 = x_1 = 0\\ \hat{x}_2 = x_2 = 0}}.
\end{align}
However, we also observe that 
\begin{align}
P_{\text{succ},4}^{\text{BPQM}} \bigg\vert_{\substack{\hat{x}_1 = x_1 = 0\\ \hat{x}_2 = x_2 = 0}} = P_{\text{succ},4}^{\text{Hel}} \bigg\vert_{\substack{\hat{x}_1 = x_1 = 0\\ \hat{x}_2 = x_2 = 0}} \neq P_{\text{succ},2}^{\text{Hel}} \bigg\vert_{\hat{x}_1 = x_1 = 0} = P_{\text{succ},2}^{\text{BPQM}} \bigg\vert_{\hat{x}_1 = x_1 = 0} = P_{\text{succ},4}^{\text{BPQM}} \bigg\vert_{\hat{x}_1 = x_1 = 0} = P_{\text{succ},4}^{\text{Hel}} \bigg\vert_{\hat{x}_1 = x_1 = 0}.
\end{align}



Therefore, overall the BPQM success probability is given by
\begin{align}
P_{\text{succ}}^{\text{BPQM}} = \mathbb{P}[\hat{\vecnot{x}} = \vecnot{x}] & = \mathbb{P}[\hat{x}_1 = x_1] \cdot \mathbb{P}[\hat{x}_2 = x_2\, \vert\, \hat{x}_1 = x_1 ] \cdot \mathbb{P}[\hat{x}_4 = x_4\, \vert\, \hat{x}_1 = x_1, \hat{x}_2 = x_2 ] \\
  & = P_{\text{succ},1}^{\text{BPQM}} \cdot P_{\text{succ},2}^{\text{BPQM}} \bigg\vert_{\hat{x}_1 = x_1 = 0} \cdot P_{\text{succ},4}^{\text{BPQM}} \bigg\vert_{\substack{\hat{x}_1 = x_1 = 0\\ \hat{x}_2 = x_2 = 0}} \\
\label{eq:Psucc_bpqm}
  & = P_{\text{succ},1}^{\text{Hel}} \left( \frac{1}{2} + \frac{1}{4 P_{\text{succ},1}^{\text{Hel}}} \norm{A \left[ \tilde{\Lambda}_2^{(1)} - \tilde{\Lambda}_2^{(0)} \right] A^{\dagger}}_1 \right) \left( \frac{1}{2} + \frac{1}{4} \norm{ \rho_{4,3}^{(000)} - \rho_{4,3}^{(001)} }_1 \right) \\
  & \neq P_{\text{succ},1}^{\text{Hel}} \left( \frac{1}{2} + \frac{1}{4 P_{\text{succ},1}^{\text{Hel}}} \norm{A \left[ \tilde{\Lambda}_2^{(1)} - \tilde{\Lambda}_2^{(0)} \right] A^{\dagger}}_1 \right)^2.
\end{align}
This success probability exactly equals the value from the closed-form expression one obtains using the fact that the square root measurement (SRM) is optimal for channel coding over the pure-state channel~\cite{Eldar-it00,Rengaswamy-arxiv20b}:
\begin{align}
\label{eq:Psucc_srm}
P_{\text{succ}}^{\text{SRM}} & = \left( \sum_{h \in \mathbb{Z}_2^k} \sqrt{\frac{\hat{s}(h)}{2^{k/2}}} \sqrt{\frac{1}{2^k}} \right)^2, \\
\label{eq:shat_final}
\frac{1}{2^{k/2}} \hat{s}(h) & \coloneqq \sum_{z \in y_h\, \oplus\, \MCC^{\perp}} p^{w_H(z)} (1 - p)^{w_H(z)}\ ; \ \ \sum_{h \in \mathbb{Z}_2^k} \frac{\hat{s}(h)}{2^{k/2}} = 1,
\end{align}
where $y_h$ is any vector in the coset of $\MCC^{\perp}$ corresponding to $h \in \mathbb{Z}_2^k$.
Alternatively, one can also use the Yuen-Kennedy-Lax (YKL) conditions~\cite{Yuen-it75,Krovi-pra15} to derive the optimal error rates.

For example, let us pick $\theta = 0.05 \mathrm{\pi}$ which corresponds to the mean photon number per mode $N \approx 0.00619$. 
Then the optimal error probability from the SRM-based closed-form expression is $0.758171401618323$ up to numerical precision.
Similarly, the density-matrix based expression~\eqref{eq:Psucc_bpqm} produces the number $0.758171401618325$ whose small difference can be attributed to numerical error.
Furthermore, we have
\begin{align}
P_{\text{succ},1}^{\text{BPQM}} \approx 0.5889, \quad 
P_{\text{succ},2}^{\text{BPQM}} \bigg\vert_{\hat{x}_1 = x_1 = 0} \approx 0.6425, \quad 
P_{\text{succ},4}^{\text{BPQM}} \bigg\vert_{\substack{\hat{x}_1 = x_1 = 0\\ \hat{x}_2 = x_2 = 0}} \approx 0.6390. 
\end{align}
Note that $P_{\text{succ},2}^{\text{BPQM}} \bigg\vert_{\hat{x}_1 = x_1 = 0} = P_{\text{succ},4}^{\text{BPQM}} \bigg\vert_{\hat{x}_1 = x_1 = 0}$ but the additional conditioning on $\hat{x}_2 = x_2$ makes a difference for $x_4$.
The overall bit error probabilities for the $5$ bits are given by
\begin{align}
P_{\text{err},1}^{\text{BPQM}} = 1 - P_{\text{succ},1}^{\text{BPQM}} \approx 0.4111, \quad 
P_{\text{err},i}^{\text{BPQM}} \approx 0.4160, 
i \in \{2,3,4,5\}.
\end{align}

To check simulation results averaged over $B = 10^6$ codeword transmissions, we set the confidence level to be $1 - \alpha = 0.98$ and calculate the accuracy $\beta$ of the error estimate.
These quantities are related as 
\begin{align}
B = \frac{1}{p} \left( \frac{Q^{-1}(\alpha/2)}{\beta} \right)^2, 
\end{align}
where $Q(\cdot)$ is the ``Q function'' of the Gaussian distribution and $p$ is the true error probability we are trying to estimate (numerically).
\begin{enumerate}

\item For the block error rate, $p \approx  0.7582$, 
we obtain $\beta \approx 0.2671\%$ which means the answer is in the window $[ 0.7561,\, 0.7602 ]$. 
The simulation produced the value $0.7573$ which is well within this window. 
When we used only $B = 10^5$ codeword transmissions we obtained the value $0.7558$. 
For this setting, again with $98\%$ confidence, the window for $\beta \approx 0.8448\%$ is $[ 0.7518,\, 0.7646 ]$, so the simulation result is well within this window. 

\item For $x_1$, the result is well within $\beta \approx 0.3629\%$ from the actual number $p \approx 0.4111$ 
since the window is $[ 0.4096,\, 0.4126 ]$ 
and the simulation gives $0.4111$. 

\item For $x_2$ through $x_5$ (which all have the same overall error probability), the results are well within $\beta \approx 0.3606\%$ from the actual number $p = 0.4160$ 
since the window is $[ 0.4145,\, 0.4175 ]$ 
and the simulation yields $0.4163$ 
for $x_2$, $0.4168$ 
for $x_3$, $0.4150$ 
for $x_4$, and $0.4163$ 
for $x_5$.

\end{enumerate}

\begin{remark}
\label{rem:coherent_rotation}
We also observe that if we ignore the coherent rotation after measuring $x_1$, then the success probabilities of the remaining bits decrease significantly to
\begin{align*}
P_{\text{succ},2}^{\text{BPQM}} \bigg\vert_{\hat{x}_1 = x_1 = 0} \approx 0.6090, 
P_{\text{succ},4}^{\text{BPQM}} \bigg\vert_{\substack{\hat{x}_1 = x_1 = 0\\ \hat{x}_2 = x_2 = 0}} \approx 0.6161. 
\end{align*}
Due to this, the overall block error rate increases to roughly $0.7790$. 
Therefore, it is clear that the coherent rotation plays an important and non-trivial role in the optimality of BPQM (for this code).
\end{remark}

\begin{remark}
\normalfont
The above analyses demonstrate that even though the measurement for each bit is irreversible, BPQM still decides each bit optimally in this $5$-bit example code.
In particular, the order in which the bits are decoded does not seem to affect the performance.
This needs to be studied further and we need to analyze if BPQM always achieves the codeword Helstrom limit for all codes with tree factor graphs.
We emphasize that, while in classical BP there is no question of ordering and one makes hard decisions on all the bits simultaneously after several BP iterations, it appears that quantum BP always has a sequential nature due to the unitarity of operations and the no-cloning theorem.
This resembles ``successive-cancellation'' type decoders more than BP.
Due to these facts, we expect that extending classical ideas for analyzing BP, such as density evolution~\cite{RU-2008}, will require some caveats in the quantum setting.


In connection to this, Renes has recently developed a precise notion of duality between channels, and shown that classical channels need to be embedded in CQ channels in order to define their duals~\cite{Renes-arxiv17}.
An interesting fact that follows from this framework is that the dual of the pure-state channel is the classical binary symmetric channel (BSC).
Since we know that density evolution is a well-defined analysis technique for BP on BSCs, albeit sophisticated, it will be interesting to see if duality allows one to borrow from this literature and analyze BPQM on pure-state channels.
\end{remark}

    
    
    

\section*{Data Availability}

No specific public/private dataset was used in this research.

\section*{Code Availability}

The computer programs to generate the data produced in our simulations are made available at \url{https://github.com/nrenga/bpqm}.

\acknowledgments

The authors acknowledge helpful discussions with Prof. Bane Vasic, Prof. Mark Neifeld, Prof. Iman Marvian, Kevin Stubbs, Sarah Brandsen, and Nithin Raveendran.
The authors would like to thank Dr. Zachary Dutton for sharing his code to numerically evaluate the YKL limit of decoding a general binary linear code~\cite{Krovi-pra15}.
The authors would also like to thank the reviewers for helpful feedback, in particular for the suggestion to compute the mutual information per photon for BPQM.

KPS and SG acknowledge the support of a National Science Foundation (NSF) project ``CIF: Medium: Iterative Quantum LDPC Decoders", award number: 1855879, and the Office of Naval Research (ONR) MURI program on Optical Computing, grant number N00014-14-1-0505. 
The work of NR and HP was supported in part by the National Science Foundation (NSF) under Grant No. 1718494, 1908730 and 1910571.  
Any opinions, findings, conclusions, and recommendations expressed in this material are those of the authors and do not necessarily reflect the views of these sponsors.

\section*{Author Contributions}

All authors were actively involved in the discussions leading to and during this research.
Specifically, Henry Pfister initiated the study of Ref.~\cite{Renes-arxiv17} which lead to analyzing the BPQM algorithm presented in Ref.~\cite{Renes-njp17}.
Narayanan Rengaswamy conducted the detailed analysis of the BPQM algorithm and the specific interpretation of it via the classical BP algorithm, aided by continuous discussions with Henry Pfister.
Saikat Guha and Kaushik Seshadreesan primarily contributed to the optical communications aspect of this work, in particular to the calculation of the Yuen-Kennedy-Lax (YKL) limit. 
They also provided the insight that the BPQM reversal after the coherent rotation following measurement of the first bit is imperfect.
They were continuously involved in discussions about the algorithm and its interpretation.

\section*{Competing Interests}

The authors declare no competing interests.

\end{document}